\documentclass[lettersize,journal,review ]{IEEEtran}
\usepackage{algorithmic}
\usepackage{algorithm}
\usepackage{array}
\usepackage[caption=false,font=normalsize,labelfont=sf,textfont=sf]{subfig}
\usepackage{textcomp}
\usepackage{stfloats}
\usepackage{url}
\usepackage{graphicx}
\usepackage{cite}
\usepackage{amsmath,amssymb,amsfonts}
\usepackage[dvipsnames]{xcolor}
\usepackage{enumitem}
\usepackage{listings}
\usepackage{color}
\usepackage{fancybox}
\usepackage{multirow}
\usepackage{hhline}
\usepackage{amsthm} 
\usepackage{tcolorbox}
\usepackage{colortbl}
\usepackage{makecell,rotating}
\usepackage{threeparttable}
\usepackage[export]{adjustbox}
\usepackage{float} 
\usepackage[misc]{ifsym}
\usepackage[hidelinks]{hyperref}
\usepackage{pifont}
\usepackage{easyReview}
\usepackage{caption} 
\usepackage[framemethod=TikZ]{mdframed}

\definecolor{mygray}{gray}{.9}
\definecolor{mycyan}{cmyk}{.3,0,0,0}
\definecolor{mauve}{rgb}{0.58,0,0.82}
\definecolor{dkgreen}{rgb}{0,0.6, 0}
\definecolor{gray}{rgb}{0.5,0.5,0.5}
\newcommand{\tabincell}[2]{\begin{tabular}{@{}#1@{}}#2\end{tabular}}
\newcommand{\ourtool}[1]{CrashTracker}
\newcommand{\journalColor}[1]{\color{black} }
\newcommand{\summaryName}[1]{ETS}

\newif\ifdraft\drafttrue
\ifdraft
\newcommand{\journal}[1]{\color{black}{#1}\color{black}}

\else
\newcommand{\journal}[1]{}
\fi
\newcolumntype{C}{c@{\hspace{0.08cm}}}
\newcolumntype{L}{l@{\hspace{0.08cm}}}

\setenumerate[1]{itemsep=0pt,partopsep=0pt,parsep=\parskip,topsep=1pt}
\setitemize[1]{itemsep=0pt,partopsep=0pt,parsep=\parskip,topsep=1pt}
\setdescription{itemsep=0pt,partopsep=0pt,parsep=\parskip,topsep=1pt}

\begin{document}

\newcommand{\prompt}[1]{
  \vspace{0.5mm}
 \begin{mdframed}[linecolor=gray,roundcorner=12pt,backgroundcolor=gray!15,linewidth=3pt,innerleftmargin=2pt,innertopmargin=2pt, leftmargin=0cm,rightmargin=0cm,topline=false,bottomline=false,rightline = false]
  #1
 \end{mdframed}
 \vspace{0.5mm}
}

\newcommand{\explanation}[1]{
  \vspace{0.5mm}
 \begin{mdframed}[linecolor=gray,roundcorner=2pt,backgroundcolor=gray!15,linewidth=1pt,innerleftmargin=3pt,innertopmargin=2pt, leftmargin=0cm,rightmargin=0cm,topline=true,bottomline=true, rightline = true]
  #1
 \end{mdframed}
 \vspace{0.5mm}
}

\title{\journal{Better Debugging: Combining Static Analysis and LLMs for Explainable Crashing Fault Localization}}

\author{Jiwei Yan, 
Jinhao Huang, 
Chunrong Fang,
Jun Yan*,
and Jian Zhang

\thanks{\textbullet  \textit{Jiwei Yan, Jinhao Huang, and Jun Yan are with the Technology Center of Software Engineering, Institute of Software, Chinese Academy of Sciences, Beijing, China. E-mail: \{yanjiwei, huangjinhao\}@otcaix.iscas.ac.cn.}}
\thanks{\textbullet \textit{Jun Yan and Jian Zhang are with the Key Laboratory of System Software, Institute of Software, Chinese Academy of Sciences, Beijing, China, and State Key Laboratory of Computer Science, Institute of Software, Chinese Academy of Sciences, Beijing, China. E-mail: \{yanjun, zj\}@ios.ac.cn.}}
\thanks{\textbullet \textit{Chunrong Fang is with the State Key Laboratory for Novel Software Technology, Nanjing University, Nanjing, China, and also with the Software Institute, Nanjing University, Nanjing, Jiangsu 210008, China. E-mail: fangchunrong@nju.edu.cn.}}
\thanks{\textbullet \textit{Jun Yan is the corresponding author.}}
}






\markboth{Journal of \LaTeX\ Class Files,~Vol.~14, No.~8, August~2021}%
{Shell \MakeLowercase{\textit{et al.}}: A Sample Article Using IEEEtran.cls for IEEE Journals}


\maketitle

\begin{abstract}
Nowadays, many applications do not exist independently but rely on various frameworks or libraries. The frequent evolution and the complex implementation of framework APIs induce many unexpected post-release crashes. Starting from the crash stack traces, existing approaches either perform direct call graph (CG) tracing or construct datasets with similar crash-fixing records to locate buggy methods. However, these approaches are limited by the completeness of CG or dependent on historical fixing records. Moreover, they fail to explain the buggy candidates by revealing their relationship with the crashing point, which decreases the efficiency of user debugging. 

To fill the gap, we propose an explainable crashing fault localization approach by combining static analysis and LLM techniques. Our primary insight is that understanding the semantics of exception-throwing statements in the framework code can help find and apprehend the buggy methods in the application code. Based on this idea, first, we design the \textit{exception-thrown summary} (ETS) that describes the key elements related to each framework-specific exception and extract ETSs by performing static analysis.
As each crash can map to a target ETS, we make data-tracking of its key elements to identify and sort buggy candidates for the given crash. Then, we introduce LLMs to improve the explainability of the localization results. To construct effective LLM prompts, we design the \textit{candidate information summary} (CIS) that describes multiple types of explanation-related contexts and then extract CISs via static analysis. Compared to SOTA fault localization works, our approach does not solely depend on CG tracing and does not require prior knowledge. Instead, it fully utilizes the information from the framework code and is the first to consider the explainability of the localization results. 
Finally, we apply our approach to one typical scenario, i.e., locating Android framework-specific crashing faults, and implement a tool called CrashTracker. For fault localization, CrashTracker exhibited an overall MRR value of 0.91 and outperformed the SOTA tool Anchor in precision. For fault explanation, compared to the naive one produced by static analysis only, the LLM-powered explanation achieved a 67.04\% improvement in users' satisfaction score.
\end{abstract}
 

\section{Introduction}

Nowadays, many applications are developed based on specific frameworks or libraries, e.g., the Android framework~\cite{AOSP}, google-map SDK~\cite{google-map}, zxing library~\cite{zxing}, etc.
When applications misuse their APIs, the framework and library-specific exceptions will be triggered, which account for the majority of app crashes and have longer fixing duration than application-specific ones~\cite{DBLP:conf/icse/FanSCMLXPS18}.
To ensure code quality, developers must thoroughly analyze crash reports to identify the buggy methods based on the report information.
Typically, the crash report includes the crash stack trace information, which shows the methods that were executed and unfinished when the crash occurred. This information is useful for developers during the code debugging process~\cite{DBLP:conf/msr/SchroterBP10}.
However, only using stack trace information is insufficient for precise fault localization.
The real buggy method may not be the last executed one in the stack trace, or even not appear in the stack~\cite{DBLP:journals/jss/GuXZZFXQ19}. Instead, after the buggy method is executed, its execution results affect the subsequent code execution and finally lead to a crash. 
Besides, considering the scale of framework-based projects, developers may have difficulty understanding the method call ordering between the application and framework code~\cite{DBLP:conf/icsm/CokerWGBS19}, which makes the debugging process difficult and costly.

To quickly fix bugs with execution results, several automatic fault localization approaches have been proposed.
The \textbf{spectrum-based methods}~\cite{DBLP:conf/pldi/LiblitNZAJ05, DBLP:conf/sigsoft/LiuYFHM05, DBLP:journals/tse/WenCTWHHC21, DBLP:journals/tosem/XieCKX13, DBLP:conf/issta/LiLZZ19,DBLP:conf/sigsoft/LouZDLSHZZ21} rank suspicious statements relying on the ratio of failed and passed test cases that execute the statements.
These techniques impose high requirements on test cases, which are not suited for post-release crashes. For these crashes, both the test cases and the runtime coverage are usually unknown.
Recent years, \textbf{learning-based approaches}~\cite{DBLP:journals/jss/GuXZZFXQ19,DBLP:journals/jss/0003ZYYXLZ20,DBLP:conf/icdm/Wang0THLXL18} are widely adopted to solve this problem, for which higher precision relies on spending more effort on dataset labeling. 
Despite working well when similar crash-fixing records are collected, they are not good at handling unfamiliar new crash reports.
For crashing faults without a high-quality test suite and similar fixing records, adopting \textbf{static-analysis-based approaches}~\cite{DBLP:conf/icse/FanSCMLXPS18, DBLP:journals/ase/KongLGRZBK21,
DBLP:conf/issta/SinhaSGJKH09,DBLP:conf/qrs/GinelliRMM21} is a more general choice, which usually tries to recover the real execution trace by backward tracking on the CG and identify the buggy one. 
\journal{Though the static-analysis-based approaches do not need historical fixing records, they deeply rely on the precision of CG, which may be incomplete, i.e., some CG edges are lost due to asynchronous methods, system or user-defined callbacks, etc.
Moreover, static analyzers have weak explainability on their localization results, i.e., only the name of the located method and the necessary localization-required data information will be reported, which is not in a user-friendly format.
Based on these results, developers may have to make a heavy effort to understand the faulty code, especially when complex framework code is involved.

In this paper, we pick static analysis as our fundamental technique to perform explainable crashing fault localization, which faces two main challenges. 
\textbf{Challenge~\ding{182}, lack of complete models for framework-specific crashing fault localization task}.
For this task, understanding the implementation of the framework is of great importance.
If we analyze the application code directly without considering the framework code, the lack of exception-triggering information may cause a loss of precision in fault localization.
However, when we consider both the application and framework code together as the target of analysis, we have to know how to use the framework information to guide the localization process. 
Even though some researchers~\cite{DBLP:conf/qrs/GinelliRMM21,DBLP:conf/issta/SinhaSGJKH09} have improved the precision of localization by manually modeling specific exceptions, there are no pre-constructed complete models for a specific framework.
\textbf{Challenge~\ding{183}, presenting static information straightforwardly in fault localization reports is not conducive to effective debugging}.
Suppose we can successfully locate the target buggy candidates, there is another challenge in making proper explanations.
Static analyzers usually report raw or structured text information (JSON, XML, etc.), which is not practical enough for real debugging.
Though generating reports with pre-defined templates can partly help to increase the readability of reports, it is still far from a user-friendly description written in natural language.}

\begin{figure}[!t]
    \setlength{\abovecaptionskip}{5pt}
    \setlength{\belowcaptionskip}{-15pt}
    \centering
    \includegraphics[width=0.45\textwidth]{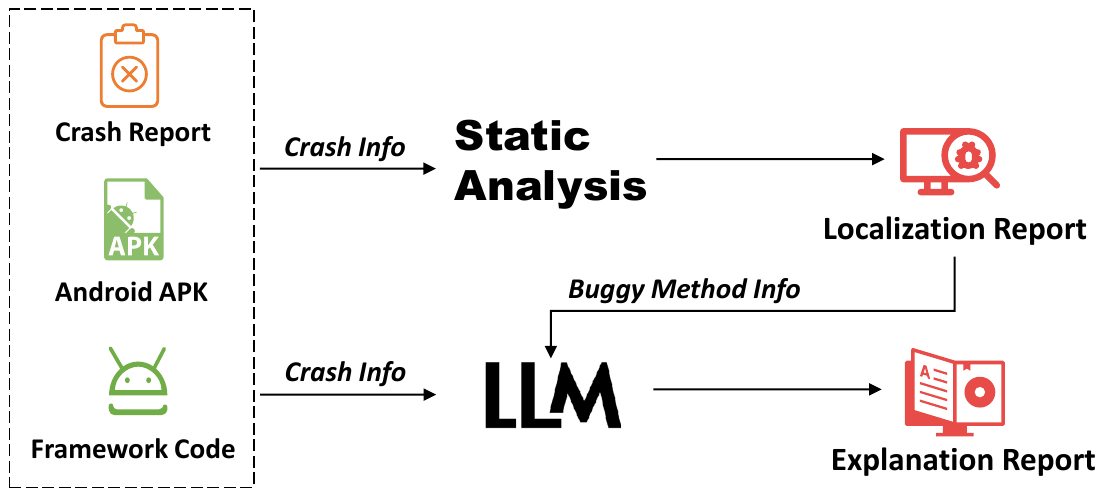}
    \caption{\journal{Workflow of Our Approach}}
    \label{fig:tool-overview-1}
\end{figure}

\journal{
In this paper, we propose a static analysis and LLM integrated explainable fault localization approach, which consists of two parts: \textbf{framework-summary-aware static fault localization} and \textbf{candidate-summary-assisting and LLM-based fault explanation}.
Fig.~\ref{fig:tool-overview-1} gives the high-level workflow of our approach. 
\textbf{To address challenge \ding{182}}, it is required to design models that fit the fault localization task and can be completely extracted.
Here, we design a model, called \textit{Exception-thrown Summary} (ETS), to describe the key fault-inducing elements that lead to exception-triggering from the framework users' point of view.
Then, we use the static analysis technique to analyze the framework-level code and automatically extract their ETS models.
After that, we analyze the application-level code to obtain the possible buggy candidates by carefully tracking the usage of key elements recorded in the crash-related ETS model.
\textbf{To address challenge \ding{183}}, we combine static analysis and LLMs, which have achieved surprising results on many software engineering tasks~\cite{DBLP:journals/pacmse/KangAY24,DBLP:conf/icse/FanGMRT23,DBLP:conf/icse/JiangLLT23,DBLP:journals/corr/abs-2308-15276}.
To provide more useful information to LLMs, we design the \textit{Candidate Information Summary} (CIS) and extract them also with static analyzers.
CISs contain all the LLM-required code snippets, the target-exception-related information, and the identified-candidate-related information.
Based on the constructed CISs, we can successfully use LLMs to generate explainable fault localization reports with high understandability.}

As Android apps are the most typical framework-based applications, we pick Android framework crashing fault localization as our evaluation task.
It has over ten million lines of code~\cite{DBLP:journals/compsec/GargB21} and millions of call edges.
Focusing on this task, we implement a tool called \textit{\ourtool{}} based on the proposed approach and perform evaluations to show its effectiveness.
First, we collect 580 instances (569 Android apps and 11 third-party SDKs) and ten versions of Android framework SDKs.
The fault localization results show that \ourtool{} exhibited an overall mean reciprocal rank (MRR) metric value of 0.91. 
It also outperforms the SOTA tool Anchor~\cite{DBLP:journals/ase/KongLGRZBK21} with 7.4\%, 11.5\%, and 12.6\% improvement on finding real buggy methods in the top 1, 5, and 10 sorted candidates.
For each crash, only 6.35 buggy candidates are provided on average. 
\journal{In the fault explanation part, the evaluation of the constructed CISs shows that it achieved high precision in context extraction. 
Specifically, the use of static analysis results helps \ourtool{} to improve the correctness of the framework constraints generation (+38.33\%).
Moreover, we perform a careful user study to evaluate the effectiveness of the LLM-enhanced reports.
Compared to the explanations generated by the original static analyzer, the updated explanations achieved a 67.04\% improvement in users' satisfaction scores.}

\textbf{Contributions.} The contributions of this work are fourfold:
\begin{itemize}[leftmargin=5pt]
\item Design \textit{exception-thrown summary} (ETS) and extract 76,247 Android framework ETSs for precise fault localization;
\item \journal{Design and build \textit{candidate information summary} (CIS) to generate LLM-powered explainable fault reports;}
\item \journal{Based on ETSs and CISs, propose an explainable fault localization approach for debugging framework-specific crashes;}
\item Implement the approach into a tool \textit{\ourtool{}} and release both the code and the experimental data publicly~\cite{CrashTracker}. 
\end{itemize}

\journal{
This paper is an extension of our previous publication in the proceedings of ICSE 2023~\cite{DBLP:conf/icse/YanWLYZ23}.  
In the journal version manuscript, we made three significant improvements.
First and foremost,
since explanations produced by plain static analysis often lack readability for real-world debugging, we first introduced LLM to generate more debugging-friendly explanations with the designed \textit{candidate information summary} (CIS) (refer to section~\ref{sec: LLM}), resulting in significantly improved results.
Though LLM has general abilities on various normal tasks, it may generate untrustworthy results on complex tasks, especially tasks that involve logical reasoning.
So, we not only use information extracted by static analysis to construct the LLM prompt but also use the static analysis results as verifiers to help filter the unreliable LLM results, which effectively improves the final result.
Second, we refined the fault localization approach with static analysis by adding illustrations for exception-triggering condition patterns, introducing the methods to identify key conditions and variables based on them, etc.
Third, we updated the prototype \ourtool{} to support generating LLM-enhanced explainable fault localization results.
To evaluate it, we conducted a series of new experiments including a user study, addressing RQ3 in section~\ref{sec: RQ3}.}

\begin{figure*}[!bp]
    \setlength{\abovecaptionskip}{5pt}
    \setlength{\belowcaptionskip}{-10pt}
    \centering
    \includegraphics[width=0.85\textwidth]{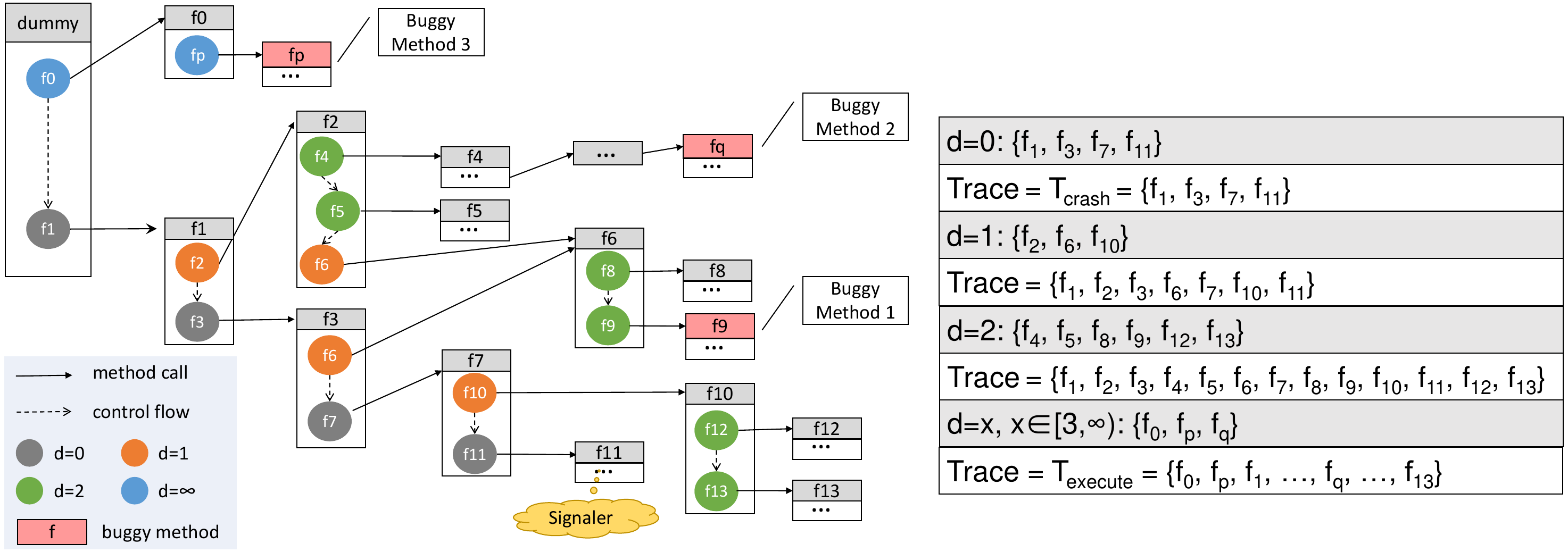}
    \caption{Execution Trace and Crash Trace}  
    \label{fig: traceExtension}
\end{figure*}

\section{Preliminary and Motivating Example}

\subsection{Application-level and Framework-level Code}
Among all the Android crashes, over 50\% are framework-specific or library-specific ones~\cite{DBLP:conf/icse/FanSCMLXPS18}, which indicates that the understanding of bottom-level code will influence code quality on the upper level.
To avoid repetitive analysis of the large-scale and complex bottom-level code, we take the methods that may need to be debugged or fixed by developers as the \textit{application-level} methods ($M_{app}$), and the methods that provide APIs to application-level methods as the \textit{framework-level} ones ($M_{frame}$).
In this paper, to debug the Android framework-specific crashes, we add all the application and library code contained in the Apks and the third-party Android SDKs into $\mathit{M_{app}}$, while adding the official Android framework code in $\mathit{M_{frame}}$.

\subsection{From Crash Report to Buggy Method} 
When a crash occurs, users usually submit crash reports to well-known platforms like GitHub \cite{Github} and StackOverflow~\cite{stackoverflow} for assistance.
The submitted crash reports mainly contain three elements, an \textit{exception type}, a \textit{crash message}, and a \textit{stack trace} snapshot that reflects the callee-caller method chain when the exception is triggered.
Table~\ref{tab:crashReport} displays the key information extracted from a crash issue report~\cite{cgeo} associated with the popular Android app \textit{cgeo}, which has 1.2k stars.
The crash report shows that an \textit{IllegalStateException} is thrown out with a crash message \textit{``attempt} \textit{to re-open $\cdots$''}. 
A crash stack trace is also given, along which we label \textit{signaler}, \textit{crashAPI}, \textit{crashMethod}, and \textit{entry} tags beside corresponding methods.
After careful debugging, the developer found the real buggy method and fixed it, which is given in the last row.
Note that, this buggy method is not in the given crash stack.

\begin{table}[!htbp] 
	\centering  
    \scriptsize
    \caption{A Crash Report and Its Real Buggy Method} \label{tab:crashReport}
	\begin{tabular}{c|l|l}  
		\hline \hline 
        \textbf{\tabincell{c}{Type}}       &\multicolumn{2}{l}{java.lang.IllegalStateException}\\ \hline
        \rowcolor{mygray}
        \textbf{Msg} &
        \multicolumn{2}{l}{\tabincell{l}{``attempt to re-open an already-closed object: SQLiteProgram: \\SELECT count(\_id) FROM cg\_caches WHERE reason $>$= 1''}} \\ \hline
        \multirow{10}{*}{\textbf{\tabincell{c}{Crash \\Stack \\Trace \\ \noindent \\ (top)\\ $\Uparrow$ \\(bottom)}}}   
                &\tabincell{l}{android.database.sqlite.SQLite-\\Closable.acquireReference} & \tabincell{l}{\textbf{signaler}:\\ the exception-thrown method}\\  \cline{2-3}
                &\tabincell{l}{android.database.sqlite.SQLite-\\Statements.simpleQuery} &\tabincell{l}{\textbf{crashAPI}:\\ the first method in M$\mathit{_{frame}}$}\\  \cline{2-3}
                &\tabincell{l}{cgeo.geocaching.DataStore\$Pr-\\eparedStmt.simpleQuery}  &\tabincell{l}{\textbf{crashMethod}:\\ the last method in M$\mathit{_{app}}$}\\  \cline{2-3}
                &\tabincell{l}{cgeo.geocaching.DataStore.get-\\AllCachesCount}  & \tabincell{l}{--}\\  \cline{2-3}
                &\tabincell{l}{cgeo.geocaching.MainActivity\$\\CountBubbleUpdateThread.run} & \tabincell{l}{\textbf{entry}:\\ the first method in M$\mathit{_{app}}$}\\ \hline
        \rowcolor{mygray}
        \textbf{Buggy}   & \multicolumn{2}{l}{cgeo.geocaching.DataStore\$PreparedStmts.clearPreparedStmts}\\ \hline \hline 
	\end{tabular}
\end{table}

When a crash is triggered, the buggy method must be located in the execution trace, i.e., it should be a method that has been executed already.
The execution trace of application $app$ can be denoted as 
$$\mathit{T_{execute}} = \langle f_0, ..., f_i,  f_{i+1}, ..., f_n  \rangle, (0 \le i, i+1 \le n).$$
In this paper, $\mathit{T_{execute}}$ is an ordered crash-triggering execution trace, where the method $f_{i+1}$ is the next method invoked after $f_i$, and the last method $f_n$ is the method that directly throws the exception out, known as the \textit{signaler} method.
As we mainly focus on framework-specific exceptions, i.e., the exceptions thrown in the framework code~\cite{DBLP:conf/icse/FanSCMLXPS18}, we have $f_n$ $\in$ $\mathit{M_{frame}}$.
The target of our fault localization is to find out the crash-leading \textit{buggyMethod}, which must be one of the executed methods.
That means, we aim to find out the method $f_b$ in $\mathit{T_{execute}}$, $f_b$ $\in$ $\mathit{M_{app}}$, and \textit{buggyMethod} = $f_b$.

During execution, methods in $\mathit{T_{execute}}$ will be pushed into a stack $st$ in order and be popped out when finished.
The crash trace $\mathit{T_{crash}}$ records the methods in the stack $st$ when the crash is triggered, whose set of elements is a subset of the complete execution trace, i.e., $\mathit{Set(T_{crash})}$ $\subseteq$ $\mathit{Set(T_{execute}}$). We can denote $\mathit{T_{crash}}$ as a sequence
$$\mathit{T_{crash}} = \langle f_e, ..., f_{i}, f_{i+p}, ..., f_n  \rangle, (0 \le e \le i, 1 \le p \le n - i),$$
where each method in it also exists in $\mathit{T_{execute}}$. 
In the crash stack trace, the top element in stack $st$ is the \textit{signaler} method $f_n$, and the bottom one is the $\mathit{entry}$ method $f_e$. Additionally, the method $f_{i+p}$ is situated next to $f_{i}$ in the stack, but it may not be the method invoked immediately after $f_{i}$, i.e. $1 \le p \le n - i$.
Note that, method $f_e$ is not equal to $f_0$ all the time.
For Java programs that have a single entry method $entry$, we do have $f_e$ = $f_0$. 
But for event-driven Android programs that have a set of callback entries $\mathit{S_{entry}}$, which contains component lifecycle and user/system callbacks, we have $f_e$ $\in$ $\mathit{S_{entry}}$.

As we can see, the trace $\mathit{T_{crash}}$ is a slice of $\mathit{T_{execute}}$. 
The \textit{buggyMethod} must exist in the trace $\mathit{T_{execute}}$.
However, it is not determined whether the \textit{buggyMethod} is in the trace $\mathit{T_{crash}}$ or not, which makes the fault localization challenging.
Starting from $\mathit{T_{crash}}$, we aim to collect methods in $\mathit{T_{execute}}$ and find out a buggy candidate set that is compact but contains the \textit{buggyMethod}.


Fig.~\ref{fig: traceExtension} introduces a simplified CG, in which each block denotes a method, and each node in the block is a method call statement labeled with the callee.
Here, $f_0$ and $f_1$ are both entry methods. We take $\mathit{dummy}$ as a dummy entry point of the program that invokes $f_0$ and $f_1$.
Among methods, solid and dotted lines are used to represent the direct method calls and control flow edges. 
Supposing there is a crash-triggering execution that starts from the entry method $f_0$ and ends with the \textit{signaler} method $f_{11}$.
When the app crashed, only four methods $\langle f_1, f_3, f_7, f_{11} \rangle$ are stored in the method stack $st$, while others have finished their execution.
As the \textit{buggyMethod} may not be in $st$, existing approaches make an expansion of stack trace based on the call edges in CG.
Their candidate size entirely depends on the call depth setting.
The \textit{call depth} of the method $f_{i}$ with respect to a given crash trace $\mathit{T_{crash}}$ is the least number of method invocation steps from any method in $\mathit{T_{crash}}$ to $f_{i}$~\cite{DBLP:conf/issta/WuZCK14}.
In Fig.~\ref{fig: traceExtension}, the nodes with the same color have the same call depth.
With a given call depth threshold $d_t$, methods with a call depth no larger than $d_t$ will be collected.
A larger depth means a higher possibility to find out the \textit{buggyMethod}.
When $d$ = 0, we only have $\mathit{T_{crash}}$. Then we increase the depth until finding the \textit{buggyMethod}. 
Supposing that $f_9$ is the \textit{buggyMethod}, when $d$ = 1, we get 7 ineffective candidates; when $d$ = 2, around twice as many candidates will be collected, which includes $f_9$.

In this scenario, there are two special cases.
First, the \textit{buggyMethod} may be far away from the mainstream, e.g., $\textit{buggyMethod}$ = $f_{q}$.
To find it out, a larger depth brings more candidates to be reviewed and increases the difficulties in fault analysis, especially for framework-specific crashing fault analysis. 
Second, the CG may be incomplete due to the existence of callbacks, native methods, or asynchronous calls, and the buggy method can be called by these unlinked methods, e.g., $\textit{buggyMethod}$ = $f_{p}$.
So, tracing along the CG requires heavy effort, but the \textit{buggyMethod} still could be lost.
Besides the application-level CG relationship, the information hidden in the framework code should also be carefully considered.





\subsection{Motivating Examples}\label{sec:motivating}
We use the real crash report displayed in Table~\ref{tab:crashReport} as our first motivating example.
All the code snippets corresponding to this crash are shown in Fig.~\ref{fig: MotivatingExample}.
First, the application-level method $\mathtt{getAllCachesCount()}$ invokes the \textit{crashMethod} $\mathtt{simpleQueryForLong()}$. 
This \textit{crashMethod} then invokes a framework-level \textit{crashAPI} in line 8, which invokes the \textit{signaler} method in line 23 and triggers an exception in lines 34-35.
For this crash, the really buggy point (lines 11-12) is in $\mathtt{clearPreparedStmts()}$, which is not shown in the stack trace $\mathit{T_{crash}}$.
The buggy reason is that the instance of $\mathtt{PreparedStmt}$ is closed without a clear operation so that it will not be reinitialized but reused directly the next time, which finally leads to a crash.
As the entry method $\mathtt{run()}$ in $\mathit{T_{crash}}$ is asynchronous, its caller is not stored in the stack, i.e., the real callback that invokes the \textit{buggyMethod} is missed.
Thus, the methods that are called by the real entry are also lost.
However, even if the real entry method is included in the stack, the \textit{buggyMethod} is still hard to retrieve, as it is far away from the crash trace and is hidden in the large candidate set. 
Therefore, to precisely identify the candidate buggy methods, we should first figure out the characteristics of the thrown exceptions and then perform analysis based on them. 
\journal{Moreover, supposing that we have successfully identified a set of ranked buggy candidates, we should also focus on providing reasonable explanations for them to assist developers in improving debugging efficiency.}

\begin{figure}[!htpb]
\centering
\setlength{\abovecaptionskip}{5pt}
\setlength{\belowcaptionskip}{-5pt}
\shadowbox{
\begin{lstlisting}
//In Android Application
public class DataStore {
    public static int getAllCachesCount(){//caller of the crashMethod
        return (int)PreparedStmt.COUNT_ALL.simpleQueryForLong();
    }
    private static class PreparedStmt{
        public long simpleQueryForLong(){ //the crashMethod
            return getStatement().simpleQueryForLong();
        }
        private static void clearPreparedStmts(){//buggyMethod
            - for (final SQLiteStatement statement : statements) {
            -     statement.close();  } //Invoke KeyAPI_2 
            + for (final PreparedStmt preparedStmt: statements){
            +     preparedStmt.statement.close();
            +     preparedStmt.statement = null;  }
            statements.clear();
        }  
    }
}
//In Android Framework
public final class SQLiteStatement extends SQLiteProgram{
    public long simpleQueryForLong() { //crashAPI
        acquireReference(); 
    }
}
public abstract class SQLiteClosable implements Closeable{
    private int mRefCount = 1;
    public void acquireReference() { //signaler method, case1
    # public void acquireReference(int id, int count) { //case 2
    #     mRefCount += count
        //mRefCount <= 0 is the keyCond
        //mRefCount is both the condVar and the externalVar
        if (mRefCount <= 0){  
            throw new IllegalStateException("attempt to 
            re-open an already-closed object: " + this); 
        }
    }
    public void releaseReference(){  //keyAPI
        boolean refCountIsZero = false;
        refCountIsZero = --mRefCount == 0;
    }
    public void close(){  //keyAPI
        releaseReference(); 
    }
}
\end{lstlisting}
}
\caption{Motivating Example of Framework-specific Exception}
\label{fig: MotivatingExample}
\end{figure}




\section{Fault Localization with Static Analysis}
This section introduces the \textbf{framework-summary-aware static fault localization approach} used by \ourtool{}.
We will first introduce the workflow of static fault localization and then discuss its details.

\subsection{Workflow of Static Fault Localization}~\label{WorkflowLocate}
\begin{figure*}[!tbp]
    \setlength{\abovecaptionskip}{5pt}
    \setlength{\belowcaptionskip}{-5pt}
    \centering
    \includegraphics[width=0.65\textwidth]{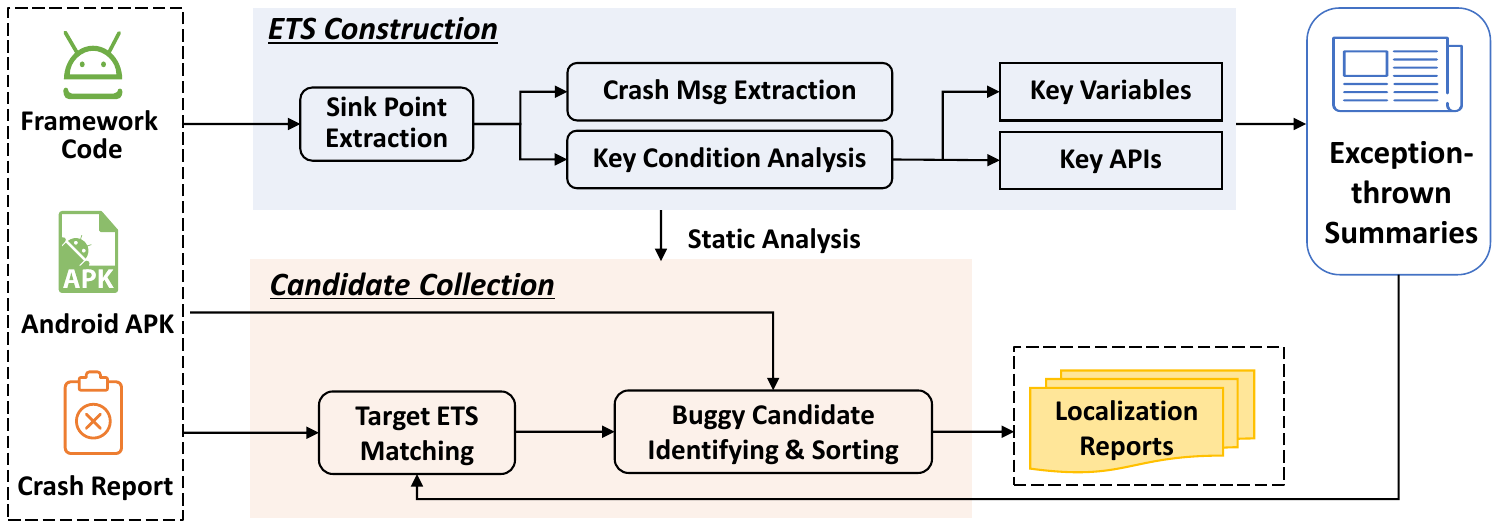}
    \caption{Overview of \ourtool{}'s Fault Localization}
    \label{fig:approach_overview}
\end{figure*}

First, we give the overview of this process in Fig.~\ref{fig:approach_overview}. It takes the framework code, application code, and crash report as input while giving a list of buggy candidates and their description texts as the original fault localization reports.
The performed analysis begins with the large-scale framework code. To avoid redundant analysis, we propose an exception-oriented summary specification that points out the fault-including elements from the view of framework users. It can be the parameters ($\mathit{keyVars}$) that are wrongly input to the framework code, or the framework APIs ($\mathit{keyAPIs}$), whose invocation can influence the status-checking results of the exception-related key conditions ($\mathit{keyConds}$). 
This specification is introduced in subsection~\ref{DefinitionofETS}.
Based on this specification, \ourtool{} performs a one-time static analysis to automatically compute summaries for framework code, which is introduced in subsection~\ref{ConstructingETS}.
In the application-level analysis, 
\ourtool{} first matches and applies the computed summary to the crash trace.
Then, according to the type of the fault-inducing elements, \ourtool{} focuses on the fault-inducing variable in the \textit{crashAPI} invocation statement and performs data tracking on it.
Also, it targets the fault-inducing APIs that are invoked in the application code and tracks its callers.
With this approach, methods that appear on the expanded CG but are not related to possible buggy data values will be excluded. Conversely, methods that are related to possible buggy data values but can not be traceable by the CG will be included.
Finally, \ourtool{} sorts all the candidates according to their buggy possibility with a heuristic strategy.
This process refers to subsection~\ref{sec:IdentifyingBuggy}.

\subsection{Definition of Exception-thrown Summary}~\label{DefinitionofETS}
The \textit{Exception-thrown Summary} (ETS) is designed to present the key fault-inducing elements for each exception-thrown point. 
It can be formally defined as a 5-tuple: 
$\mathcal{ETS}(e)$ $=\langle \mathit{id, S_{keyCond}, S_{keyCondVar}, S_{keyVar}, S_{keyAPI}}\rangle$, where
\begin{itemize}[leftmargin=10pt]
    \item $id$ is the identifier of exception $e$, which is a 4-tuple $\langle sink, signaler, type, msg \rangle$, in which $\mathit{sink}$ is a statement that throws the exception $e$; $\mathit{signaler}$ is the method that contains $\mathit{sink}$; $\mathit{type}$ is the type of the exception $e$; and $\mathit{msg}$ is the exception's description message that composed of constants and variables. When $e$ is thrown, we can get a message instance of the exception as all the variables have been dynamically assigned. Considering the existence of variables, we represent $\mathit{msg}$ by regular expressions to match message instances in all the forms; 

    \item $\mathit{S_{keyCond}}$ is a set of key conditions ($\mathit{keyCond}$ $\in$ $\mathit{S_{keyCond}}$) located in $\mathit{signaler}$, whose results can decide whether $e$ is triggered. If $e$ is triggered only when $\mathit{keyCond}$ is satisfied, $\mathit{keyCond}$ is a \textit{basic check}. 
    If the \texttt{throw($e$)} statement can not be executed as the satisfaction of  $\mathit{keyCond}$ leads to method return, $\mathit{keyCond}$ is a \textit{not-return check}; If e is triggered when another exception is caught, $\mathit{keyCond}$ is a \textit{try-catch check}.
	
    \item $\mathit{S_{keyCondVar}}$ is a set of variables $\mathit{keyCondVars}$, whose values are directly checked in $\mathit{S_{keyCond}}$;
	
    \item $\mathit{S_{keyVar}}$ denotes a set of key variables, which can influence the value of $\mathit{keyCondVar}$ and can be modified by framework users. Each $\mathit{keyVar}\in\mathit{S_{keyVar}}$ is a triple $\langle \mathit{mtd, loc, keyCondVar} \rangle$, in which $\mathit{mtd}$ is a framework-level public method; $loc$ is a parameter location in method $\mathit{mtd}$; the $loc^{th}$ parameter in $\mathit{mtd}$ can influence the value of key condition variables in $\mathit{S_{keyCondVar}}$ by inter-procedural parameter passing, i.e., its value can affect the checking results of the key conditions. For simplicity, we will use $\mathit{keyVar}$ to denote the parameter variable in $\mathit{keyVar.mtd}$ with location $loc$.

    \item $\mathit{S_{keyAPI}}$ denotes a set of key APIs, which can influence the value of $\mathit{keyCondVar}$ and can be invoked by framework users. Each $\mathit{keyAPI}\in\mathit{S_{keyAPI}}$ is a 4-tuple $\langle \mathit{mtd, keyField, keyCondVar, dpt} \rangle$, in which $\mathit{mtd}$ is a framework-level public method; $\mathit{keyField}$ is a class field that is modified by $\mathit{mtd}$ or its callees, whose value is data-related with the key condition variables in $\mathit{S_{keyCondVar}}$; $dpt$ records the least number of method invocation steps from $\mathit{mtd}$ to the method that directly modifies $\mathit{keyField}$, which is used in candidate sorting. Subsequently, we will use $\mathit{keyAPI}$ to denote the method $\mathit{keyAPI.mtd}$.
\end{itemize}

Here, we use the running example in Fig. \ref{fig: MotivatingExample} to exemplify the defined elements in ETS. 
For the exception thrown in lines 34-35, its $\mathit{keyCond}$ is $\mathtt{mRefCount\le0}$. $\mathtt{mRefCount}$ is a $\mathit{keyCondVar}$.
It is also a field of class $\mathtt{SQLiteClosable}$ that can be modified out of the $\mathit{signaler}$ method. 
As public methods $\mathtt{releaseReference()}$ and $\mathtt{close()}$ directly and indirectly modify its value,
$\langle\mathtt{releaseReference()}, \mathtt{mRefCount}, \mathtt{mRefCount}, 1\rangle$ and $\langle \mathtt{close()}, \mathtt{mRefCount}, \mathtt{mRefCount}, 2\rangle$ are added into $\mathit{S_{keyAPI}}$.
In this example, as no parameter variable is related to the $\mathit{keyCondVar}$, $\mathit{S_{keyVar}} = \emptyset$. 
But if line 28 is replaced by lines 29-30 (case 2), we will get a parameter-related variable $\mathtt{count}$, which can be modified outside and influence the value of $\mathit{keyCondVar}$.
As the method $\mathtt{acquireReference()}$  is public, we will add $\langle \mathtt{acquireReference()}, 2, \mathtt{mRefCount}\rangle$ into $\mathit{S_{keyVar}}$ and trace its callers to identify more $\mathit{keyVars}$.

\subsection{Constructing Exception-thrown Summaries}~\label{ConstructingETS}
The process of constructing ETS is displayed in Algorithm~\ref{alg: framework}, which is designed to gather the elements defined in subsection~\ref{DefinitionofETS}. The initial step involves identifying all the points in the framework code where exceptions are thrown, 
i.e., identify all the exception instances.
Starting from these instances, we perform static analysis to collect other essential ETS information, including obtaining \textit{keyConds} and \textit{keyCondVars}, collecting \textit{externalVars}, \textit{keyVars}, and \textit{keyAPIs}, etc.

\begin{algorithm}[!t]
	\caption{Exception-thrown Summary Extraction} \label{alg: framework}
	\setlength{\abovecaptionskip}{-5pt}
	\setlength{\belowcaptionskip}{-10pt}
	\small
	\begin{algorithmic}[1]
		\REQUIRE method $\mathit{signaler}$ in $\mathit{M_{frame}}$
		\ENSURE the exception summary set $S_{ets}$ on method $\mathit{signaler}$
        \FOR {sink in signaler.getSinkPoints()}
            \STATE $\mathit{ets}$ = createETS($\mathit{signaler}$, $\mathit{sink}$)
            \STATE $\mathit{ets}$.type = exceptionTypeAnalysis()
            \STATE $\mathit{ets}$.message = regexStringAnalysis()
            \STATE $\mathit{cfg}$ = $\mathit{signaler}$.getCFG()
            \STATE getKeyCondsAndVars($\mathit{ets}$, $\mathit{cfg}$, $\mathit{sink}$)
            \IF{$\mathit{ets}$.keyConds.size()$=$0}
                \FOR{return stmt $\mathit{retStmt}$ predsOf $\mathit{sink}$ in $\mathit{cfg}$}
                    \STATE getKeyCondsAndVars($\mathit{ets}$, $\mathit{cfg}$, $\mathit{retStmt}$)
                \ENDFOR    
            \ENDIF
            \STATE getKeyCondsAndVarsForTryBlock($\mathit{ets}$, getTryBlock($\mathit{sink}$))
            \STATE $\mathit{worklist}$ = $\mathit{ets}$.keyCondVars.copy()
            \WHILE {$\mathit{worklist}$.size()$>$0}
                \IF{$\mathit{worklist}$.get(0) is parameter or field related}
                    \STATE add $\mathit{worklist}$.getAndPop(0) to $\mathit{externalVars}$
                \ELSE 
                    \STATE $\mathit{defStmt}$ = getDefStmt($\mathit{worklist}$.getAndPop(0))
                    \STATE $\mathit{worklist}$.add($\mathit{defStmt}$.getRightOp().getVars())
                \ENDIF
            \ENDWHILE
            \FOR{$\mathit{var_p}$ in getParameterExternalVars($\mathit{externalVars}$)}
                \STATE add $\mathit{var_p}$ to $\mathit{ets}$.keyVars if $\mathit{signaler}$ is public
                \STATE track $\mathit{var_p}$ in $\mathit{signaler}$'s caller to update $\mathit{ets}$.keyVars
            \ENDFOR
            \FOR{$\mathit{var_f}$ in getFieldExternalVars($\mathit{externalVars}$)}
                \STATE add public methods that modify $\mathit{var_f}$ into $\mathit{ets}$.keyAPIs
                \STATE update public caller of keyAPIs into $\mathit{ets}$.keyAPIs 
            \ENDFOR
            \STATE $S_{ets}$.add($\mathit{ets}$)
		\ENDFOR
		\RETURN $S_{ets}$
	\end{algorithmic}
\end{algorithm}
\begin{algorithm}[!t]
	\caption{getKeyCondsAndVars} \label{alg: framework2}
	\small
	\begin{algorithmic}[1]
        \REQUIRE ETS $\mathit{ets}$, control flow graph $\mathit{cfg}$, statement $s$
        \ENSURE updated ETS $\mathit{ets}$
        \FOR{condition check stmt $condStmt$ predsOf $s$ in $\mathit{cfg}$}
            \STATE $\mathit{ets}$.keyConds.add($condStmt$.getCond())
            \STATE $\mathit{ets}$.keyCondVars.add($condStmt$.getCond().getVars())
        \ENDFOR
    \end{algorithmic}
\end{algorithm}

\textbf{Identifying Exception Instances.}
First, all the $\mathtt{throw(e)}$~\cite{throw} invocation points are sink points.
Besides, developers may also customize exception-thrown information and use the logged information to debug. 
To recognize them, we detect all the $\mathtt{Throwable}$ instances and trace their data flows.
For the exceptions that are not thrown directly, we take the methods that receive these instances as exception-handling methods and take the invocation statements of them as sink points.
The $\mathtt{throw(e)}$ statement is the most frequently used sink point type with many instances. 
Besides it, we get another 33 types of sink points, of which 12 store exception traces into logs or files.
Starting from the sink points, one challenge is how to extract the description message of the target exception (line 4).
For the same exception, the runtime crash message may be different, as the values of some variables are dynamically assigned.
To make a precise matching, we transform the exception message into a regular expression~\cite{regularExpr} pattern, so that it can match multiple runtime crash messages.
The method \textit{regexStringAnalysis()} in line 4 tracks the definition statement of the target exception, which may be a newly created one, e.g., $\mathtt{e = new RuntimeException()}$, or the alias of another exception, e.g., $\mathtt{e = getFileException(\cdots)}$.
For the former, we perform backward value tracing of the message-related parameter in the exception's constructor method. 
For the latter one, we make inter-procedure tracing to get the real instantiate point and analyze it as the former.
During the value tracing, we model a set of String-related APIs to stitch multiple parts together, in which we use [\verb|\|\texttt{s}\verb|\|\texttt{S}]* to represent a symbolic value and use \verb|\|\texttt{Q}$\mathtt{str}$\verb|\|\texttt{E} to represent the constant value of $\mathtt{str}$.
In lines 34-35 of Fig.~\ref{fig: MotivatingExample}, the target exception message is ``\verb|\|\texttt{Q}attempt to re-open an  
already-closed object: \verb|\|\texttt{E}[\verb|\|\texttt{s}\verb|\|\texttt{S}]*'', which can matche the crash message in Table~\ref{tab:crashReport}.


\journal{
\textbf{Obtaining KeyConds and KeyCondVars.}
Each exception is influenced by a set of conditions ($\mathit{keyConds}$), whose checking results decide whether the exception can be triggered.
Overall, three types of conditions can influence the triggering of exceptions, which are presented in listings~\ref{fig: Example1}-\ref{fig: Example5}. For these checks, listings~\ref{fig: Example1}-\ref{fig: Example3} belong to \textit{basic checks}, listings~\ref{fig: Example4} is the \textit{not-return check}, and Listing~\ref{fig: Example5} is the \textit{try-catch check}.
}

\lstset{
    basicstyle={\footnotesize\ttfamily\centering},
    frame=l,
    numbers=none,
    captionpos=b, 
    caption=\footnotesize,
    aboveskip=2pt,
    framerule=1pt,
    rulecolor=\color{dkgreen},
    backgroundcolor=\color{gray!5}
}

\journal{1) For \textit{single if} or \textit{single-switch-case} type basic check, we extract the direct single condition of the exception-thrown point. In Listing~\ref{fig: Example1}, $\mathtt{a==0}$ is the $\mathit{keyCond}$ for target exception. In the following parts, we will not list the switch-case type as a special format as it is similar to the if-else type.} 

\begin{lstlisting}[caption={\journal{Example for Single If/Switch-case Condition}}, label={fig: Example1}]
public void Example1 (int a) throws Exception {
    if(a==0) 
        throw new MyException("if condition"); //target exception
}
public void Example1(String type) throws Exception {
    switch(type)
        case "x": throw new MyException("switch condition");
}
\end{lstlisting}

\journal{2) For \textit{multiple-if} or \textit{if-and} type basic check, multiple conditions should be collected. For example, $\mathtt{type!=null}$, $\mathtt{a<0}$ and $\mathtt{b<0}$ are all $\mathit{keyConds}$ for the target exception in Listing~\ref{fig: Example2}. Sometimes, one exception may be located in a relatively deep position (about 2.5\% of the ETSs in the Android framework contain more than 10 conditions). Considering the conditions far away from the exception-thrown point may have a weak relationship with the exception, we count the average condition length as a threshold size of $\mathit{S_{keyCond}}$.}
\begin{lstlisting}[caption={\journal{Example for Multiple-if/If-and Condition}}, label={fig: Example2}]
public void Example2 (int a, int b) throws Exception {
    if(type!=null) {
        if(a<0 && b<0)
            throw new MyException("multiple if condition"); //target exception
    }
}
\end{lstlisting}

\journal{3) For \textit{if-or} type basic check, it corresponds to multiple paths in the IR level. Any path could be the real path in further execution. To guarantee the completeness of condition collection, we take all the conditions as $\mathit{keyConds}$ we are concerned about. In Listing~\ref{fig: Example3}, conditions $\mathtt{a>0}$ and $\mathtt{b>0}$ are both extracted as $\mathit{keyConds}$ for the target exception.}
\begin{lstlisting}[caption={\journal{Example for If-or Condition}}, label={fig: Example3}]
public void Example3 (int a, int b) throws Exception {
    if(a>0 || b>0)
        throw new MyException("multiple path condition"); //target exception
}
\end{lstlisting}

\journal{4) In another case, developers may take the method-return operation as the expected behavior and throw an exception if the method doesn't jump out in time.
    The conditions related to these return statements are also key conditions.
    For \textit{not-return check}, the condition for the return statement also decides whether the exception can be thrown.
    In Listing~\ref{fig: Example4}, $\mathtt{type==null}$ is the $\mathit{keyCond}$ for the target exception.}
\begin{lstlisting}[caption={\journal{Example for Not-return Condition}}, label={fig: Example4}]
public void Example4 (String type) throws Exception {
    if(type != null) return;
    throw new MyException("not-return condition"); //target exception
}
\end{lstlisting}

\journal{5) Finally, for \textit{try-catch check}, an exception is triggered when a specific type of exception is caught, so the exception triggering may be related to all the statements in the try-catch block. 
    To find the most related elements, we check the exception-thrown declaration of each invoked method to see if they can match the currently caught exception. 
    If such a statement is found, variables in that invocation statement are data-related variables, or else, other variables in the try-catch block should also be recorded.
    In Listing~\ref{fig: Example5}, the callee method $\mathtt{throwExpCall(b)}$ throws an $\mathtt{IllegalArgumentException}$, which is the subclass of $\mathtt{RuntimeException}$. As this exception will be caught, we only label variables used in statement $\mathtt{throwExpCall(b)}$ as $\mathit{keyCondVars}$. By inter-procedure analysis, $\mathtt{b>0}$ is the $\mathit{keyCond}$ for the target exception.}
\begin{lstlisting}[caption={\journal{Example for Try-catch Condition}}, label={fig: Example5}]
public void Example5 (int a, int b, String type) throws Exception {
    try {
        mormalCall(a); 
        throwExpCall(b);
    } catch (RuntimeException e) {
        throw new MyException("invalid type: " + type); //target exception
    }
}
public void throwExpCall(int p) throws IllegalArgumentException {
    if(p>0)  //throw a subclass of RuntimeException
        throw new IllegalArgumentException("illegal argument " + p);
}
\end{lstlisting}

\journal{
In Algorithm~\ref{alg: framework}, lines 6-12 are used to extract the key conditions and variables.
In line~6, we invoke method $\mathtt{getKeyCondsAndVars()}$ to trace all the predecessor statements of the sink point, i.e., exception-thrown statement, in the control flow graph (CFG), record the involved condition checks into $\mathit{S_{keyCond}}$, and collect all the condition-related variables into $\mathit{S_{keyCondVar}}$, as displayed in Algorithm~\ref{alg: framework2}.
In lines 7-11, if there is no \textit{basic check}, we extract the \textit{not-return checks} by analyzing the basic checks of all the return statements prior to the sink point in the CFG.
In line 12, the \textit{try-catch check} is handled by heuristically matching the exception type caught in the current method with the one thrown in the callee method.}

\textbf{Collecting ExternalVars, KeyVars and KeyAPIs.}
To find out $\mathit{keyVars}$ and $\mathit{keyAPIs}$, we use a worklist algorithm to locate the method parameters and class fields that influence the value of $\mathit{keyCondVars}$ by performing backward data tracing along the use-def-chains~\cite{Use-define-chain} (lines 13-21). 
Note that, in line 18, method $\mathtt{getDefStm()}$ returns the Jimple~\cite{jimple} IR level definition statements, which reflect the variables that can influence that target one.  
As these variables can be modified outside $\mathit{signaler}$, they are called external variables $\mathit{externalVars}$.
In lines 22-25, for each parameter-related $\mathit{externalVars}$ $\mathit{var_p}$, we record its method, the location in the parameter list, and the influenced $\mathit{keyCondVar}$. 
If the $\mathit{signaler}$ is a public method, $\mathit{var_p}$ itself is a $\mathit{keyVar}$ (line 23), e.g., we have $\langle\mathtt{acquireReference(), 2, mRefCount}\rangle$ for case 2 in Fig.~\ref{fig: MotivatingExample}.
Similarly, we perform backward inter-procedural parameter call-chain analysis (line 24) to trace more key variables passed through other framework APIs, which invoke the $\mathit{signaler}$.
That is, for a formal parameter, we find the actual parameter in its caller and judge whether the passed variable is influenced by the caller's parameter-related $\mathit{externalVars}$.
If it is, these newly detected external variables in the public callers will be added into $\mathit{S_{keyVar}}$, e.g., for method $\mathtt{f(int\ count)}$ $\mathtt{\{acquireReference(1,count)\}}$, we further have $\langle \mathtt{f(), 1, mRefCount} \rangle$.
In lines 26-29, for each field-related external variable $\mathit{var_f}$, its value can be changed by other framework APIs that influence the checking results of $\mathit{keyConds}$. In line 27, we record the methods that change the value of $\mathit{var_f}$, the field $\mathit{var_f}$, the influenced $\mathit{keyCondVar}$, and the call depth from the current method to the one that directly modifies
$\mathit{keyField}$. 
For Fig.~\ref{fig: MotivatingExample}, we first get the $\mathit{keyAPI}$ $\langle\mathtt{releaseReference(), mRefCount, mRefCount, 1}\rangle$. 
Then in line 28, we trace callers of the collected $\mathit{keyAPIs}$ and get $\langle\mathtt{close(), mRefCount, mRefCount, 2}\rangle$.
Finally, in line 30, we add ETS for each exception into the ETS set $S_{ets}$.

\subsection{Identifying and Sorting Buggy Candidates with ETS} ~\label{sec:IdentifyingBuggy}
Algorithm~\ref{alg:localization} displays the process of crash fault localization on $M_{app}$. 
The first step is to use the crash information to match the target ETS from multiple framework versions.
In line 1, we match the given crash message with the regular expression format message of each ETS.
When a specific framework version is given, the target ETS is unique and can be used directly.
If the version is undetermined, multiple ETSs may be matched as the exception with the same type and the description message can exist in many versions. 
In the latter case, we first classify the matched ETSs by their characteristics into five ETS-related types, which are listed in Table~\ref{tab:strategies}, and then pick the target ETS from the most classified type.

\begin{algorithm}[!htbp]
	\caption{Crash Fault Localization} \label{alg:localization}
	\setlength{\abovecaptionskip}{5pt}
	\setlength{\belowcaptionskip}{-10pt}
	\small
	\begin{algorithmic}[1]
		\REQUIRE Android app $app$, crash stack trace $st$ and message $\mathit{msg}$, framework version $\mathit{ver}$, exception summary set $S_{ets}$ on $\mathit{M_{frame}}$
		\ENSURE Buggy candidates $\mathit{candis}$
        \STATE $\mathit{ets}$ = getBestMatchETS($st$, $\mathit{msg}$, $S_{ets}$, $\mathit{ver}$)
        \STATE $\mathit{ETSRelatedType}$ = getETSRelatedType($\mathit{ets}$)
        \STATE $\mathit{strategies}$ = getStrategiesByType($\mathit{ETSRelatedType}$)
        \FOR {$\mathit{strategy}$ in $\mathit{strategies}$}
            \IF{$\mathit{strategy}$ = $S_1$} 
                \STATE locate methods that should override $ets.signaler$
            \ELSIF{$\mathit{strategy}$ = $S_{2}$}
                \STATE locate methods that are data-related with the crashAPI-invoking statement
            \ELSIF{$\mathit{strategy}$ = $S_3$}
                \STATE locate methods that are data-related with $keyVars$
            \ELSIF{$\mathit{strategy}$ = $S_4$}
                \STATE locate methods that are callers of $keyAPIs$
            \ENDIF
        \ENDFOR
        \STATE $\mathit{candis}$ = filterAndSortCandidates()
		\RETURN $\mathit{candis}$
	\end{algorithmic}
\end{algorithm}

In lines 2-14 of Algorithm~\ref{alg:localization}, according to the information provided by the target ETS, we use different candidate-picking strategies. 
Table~\ref{tab:strategies} also displays the four strategies that can be used for each ETS-related type.

\begin{itemize}[leftmargin=10pt]
\item For the ETS who has \textit{no condition variable}, i.e., no $\mathit{keyCondVar}$ and $\mathit{keyCond}$, we use strategy $S_1$.
In this case, the $\mathit{signaler}$ method should not be invoked directly, framework users should override that method and invoke the newly implemented method by the polymorphic mechanism. 

\item For target ETS who has $\mathit{keyCondVar}$ but has \textit{no external variable}, we use strategy $S_2$.
The exceptions the exceptions whose $\mathit{keyCondVars}$ are related to methods with unknown implementations (e.g., native method) is in this type.
For this type, we will make data tracing starting from the invocation statement of \textit{crashAPI} in the \textit{crashMethod}.

\item The strategy $S_{3}$ is for ETSs that \textit{only have keyVars}.
The difference with strategy $S_2$ is that, first, we can confirm this crash is caused by the wrong parameter value.
And we may get the location of the fault-inducing parameters. So that, we can focus on where these target parameters are created or assigned, and perform a target-directly data tracing.

\item Similarly, the strategy $S_{4}$ targets ETSs that \textit{only have keyAPIs}.
These crashes have no relation to the passed parameter but are influenced by the previously invoked APIs, which change the value of class fields in the framework code and further influence the exception's condition-checking results. To cope with them, we focus on the methods that invoke the $\mathit{keyAPIs}$ and further track their callers.
In the motivating example, even if method $\mathtt{clearPreparedStmts()}$ cannot be traced along the CG, we still can find it as it invokes the $\mathit{keyAPI}$ $\mathtt{close()}$.

\item Finally, one ETS can \textit{have both keyVars \& keyAPIs}.
In this case, we collect all the methods tracked by either strategies $S_{3}$ or $S_{4}$ as possible candidates.
\end{itemize}

\begin{table}[htbp] 
\centering  
\setlength{\abovecaptionskip}{5pt}
\setlength{\belowcaptionskip}{-5pt}
\caption{ETS-related Types \& Fault Localization Strategies }\label{tab:strategies}
\begin{tabular}{L|l}  
    \hline \hline 
    \rowcolor{mygray}
    \textbf{ETS-related Types}& \textbf{Fault Localization Strategy} \\ \Xhline{1pt}
    No keyCondVar &$S_1$: Override analysis in subclasses\\ \hline
    No externalVar     &$S_{2}$: Data tracing from variables in crashAPI$_{inv}$\\ \hline
    Only keyVar     &$S_{3}$: Data tracing of variables \textit{keyVars}\\ \hline
    Only keyAPI     &$S_4$: Call tracing of methods invoking \textit{keyAPIs} \\ \hline
    Both keyVar \& keyAPI & $S_3$ + $S_4$ \\\hline\hline 
\end{tabular}  
 \vspace{-2mm}
\end{table}

To make the candidate set in a compact size, we perform universal filtering and adjustment on all the identified candidates.
First, we extract the \textit{package} and \textit{class} characteristics of candidates.
At the package level, we filter the candidates that have different package prefixes (e.g., first two elements) with all the methods in $T_{crash}$, as these methods are too far away from the \textit{crashMethod} to be the right buggy method.
As library methods can also be traced through control- and data-flow tracing, we make a penalty on the methods that are not in the app-declared package. This penalty helps to decrease their priority. 
Users can customize it if the specific packages should be considered with high priority in the fault localization. 
Specifically, we observed that many methods work as utility functions and have many callers.
One heuristic strategy is to stop tracing the caller methods when the number of callers exceeds the user-defined upper limit.
In our implementation, when tracing the application-level callers of $\mathit{keyAPIs}$, the limit is set as ten according to our experience.
Finally, as a supplement, the methods in $T_{crash}$ are added by default with a conservative score.

The candidates are mainly collected by data tracing from a set of variables or call tracing from specific APIs.
In line 15, for the candidate $f_i$ collected by the data-tracing of variables, we measure its distance to the \textit{crashMethod} by the formula
$\mathit{dis(f_i) = min (callDepth(f_i, f_j) + callDepth(f_j, f_{cm}))}$,
where $f_j$ is a method located in $\mathit{T_{crash}}$ and $f_{cm}$ is the \textit{crashMethod}.
For each candidate, we have an initial score of $init$. The longer the distance, the larger the score penalty.
Their scores are computed by $\mathit{score(f_i) = init - dis(f_i)}$.
And for the candidates collected by the call-tracing from specific APIs, if method $f_k$ invokes the $\mathit{keyAPI}$ $api$, its score can be computed by $\mathit{score(f_k) = init - api.dpt}$, in which $dpt$ is the least distance from $\mathit{keyAPI}$ to the field manipulating method.
If $f_t$ is a caller of $f_k$, its score can be computed by $\mathit{score(f_t) = init - callDepth(f_t, f_k) - api.dpt}$.

\journal{
\section{Fault Explanation with Static Analysis and LLM Integration} \label{sec: LLM}
This section introduces the \textbf{candidate-summary-assisting and LLM-based fault explanation} used by \ourtool{}.
Although static analysis can help us precisely locate buggy candidates, users are still struggling to figure out the relationship between buggy candidates and the crash points.
Faced with this problem, we aim to generate fine-grained fault explanation reports based on novel LLMs to make the debugging process easier. This section first introduces the workflow of LLM-powered fault explanation and then discusses its details.

\subsection{Workflow of LLM-based Explanation in \ourtool{}}~\label{WorkflowExplain}
Fig.~\ref{fig: WorkflowOurLLM} gives the overview of the LLM-based explanation process, which takes the CrashTracker's static analysis results as input while giving the final explanation report as output.
After fault localization with static analysis, we can get the target ETS, the buggy candidate set, as well as a templated naive explanation report.
Based on that, in the explanation generation phase, our key insight is to sufficiently use these static analysis results to assist LLM in generating debugging-friendly explanations.
Here, we design three types of information collectors to obtain the LLM-required code snippets, the application-level candidate-related information as well as the framework-level exception-related constraint information. All of them form the candidate information summaries (CISs), which are passed to LLM for explanation generation.
In subsections~\ref{LLMSummaryD} and~\ref{LLMSummaryC}, we first discuss the definition and the construction of CISs.
Then, subsection~\ref{LLMExplain} introduces how to dynamically construct proper LLM prompts and generate the final explanation reports with CISs.

\begin{figure*}[!htbp]
    \setlength{\abovecaptionskip}{5pt}
    \setlength{\belowcaptionskip}{-5pt}
    \centering
    \includegraphics[width=0.88\textwidth]{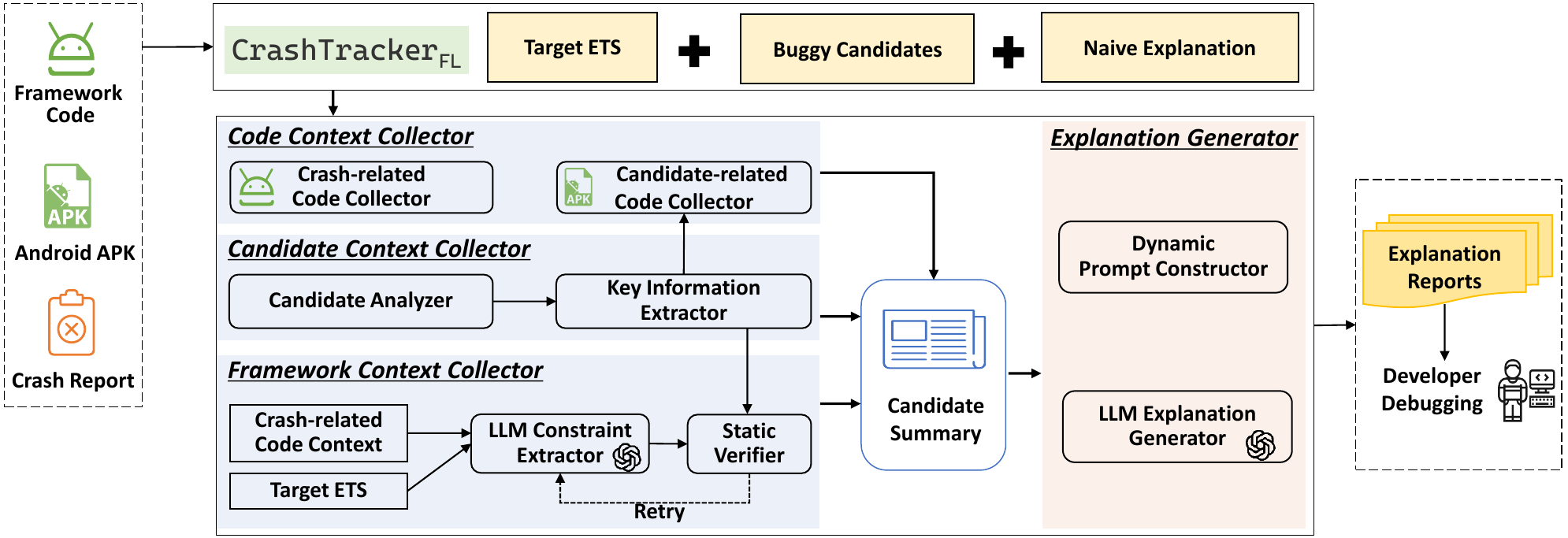}
    \caption{Workflow of \ourtool{}'s Fault Explanation}
    \label{fig: WorkflowOurLLM}
    \vspace{-2mm}
\end{figure*}

\subsection{Definition of Candidate Information Summary}~\label{LLMSummaryD}
The \textit{Candidate Information Summary} (CIS) is designed to provide sufficient context for LLM explanation. 
For a buggy candidate method $b$ of the crash $c$, its CIS can be formally defined as 
$\mathcal{CIS}(b,c)=\langle id, code, \mathit{keyInfo}, \mathit{constraint}\rangle$, where
\begin{itemize}[leftmargin=10pt]
    \item $id$ is the id of the summary, which is a tuple $\langle b, c \rangle$;
    \item $code$ denotes the \textit{code context}, including the buggy-candidate-related code (in $\mathit{M_{app}}$) that will be used as LLM context, and the crash-triggering-related code (in $\mathit{M_{frame}}$) that will be used in framework context extraction.
    \item $\mathit{keyInfo}$ denotes the \textit{candidate context}, i.e., the key information about the buggy candidate method $b$, including the pattern of a candidate in terms of its identification, and the key elements involved under that pattern. 
    \item $\mathit{constraint}$ denotes the \textit{framework context}, i.e., the constraints that should be satisfied before calling a framework API to avoid triggering a specific exception.
    For a crash report, we only concern the constraints for its \textit{crashAPI};    
\end{itemize}

\subsection{Constructing Candidate Information Summaries}~\label{LLMSummaryC}
The following parts will introduce how the key parts in CISs are extracted and used.

\subsubsection{\textbf{Collecting Code Context}}
First, we collect all the executed methods in the crash stack trace as code context.
To achieve that, we retrieve both the Android Open Source Project (AOSP)~\cite{AOSP} and the app's source code.
If the app's source code is unavailable, we use Jadx class decompiler~\cite{Jadx} to get the decompiled source code and retrieve code from it.
However, when only stack trace code snippets and the buggy candidate method are given to LLM, it may fail to understand the real buggy scenario and produce hallucinations.
To cope with this, for the crash-triggering-related code in $\mathit{M_{frame}}$, we also retrieve the methods related to the target exception according to the ETS information;
and for the buggy-candidate-related code in $M_{app}$, we retrieve methods associated with the buggy candidate based on the recorded key information, as described in subsection~\ref{ObtainLLMInfo}.


\subsubsection{\textbf{Collecting Candidate Context}}~\label{ObtainLLMInfo}
To clearly describe why a candidate may be the buggy method, more specific characteristics for each candidate are collected as candidate context.
\ourtool{} first categorizes candidates into different \textit{Explanation Patterns} (EPs) according to how the static analyzer identifies these candidates. Then it extracts the key
elements involved in the given EP. The pattern and the key elements are collectively referred to as the key information, i.e., \textit{CIS.keyInfo}.
There are four EPs displayed as follows.

\textbf{EP 1: Invoking unconditionally-thrown exception.} 
This EP denotes that the given candidate invokes a framework method throwing the target exception without any condition check.
As shown in Listing~\ref{fig: EP1}, the method $\mathtt{noCondThrow()}$ in the framework class $\mathtt{F1}$ throws an exception unconditionally.
The application class $\mathtt{A1}$ extends the framework class $\mathtt{F1}$. However, it does not override the method $\mathtt{noCondThrow()}$.
Therefore, whenever $\mathtt{noCondThrow()}$ is invoked by the instance of $\mathtt{A1}$, an exception will be thrown.
}
\begin{lstlisting}[caption={\journal{Invoking Unconditionally-thrown Exception}}, label={fig: EP1}]
public class A1 extends F1{  
    //noCondThrow is not overwritten but invoked  
}

public class F1 {
    public void noCondThrow(){
        throw new SomeException(); //throw exception without any condition
    } 
}
\end{lstlisting}

\journal{
The key elements to be collected for this EP are as follows: 
\begin{itemize}[leftmargin=10pt]
    \item Signaler: the signature of the last invoked framework method in the crash stack, which is the one to be overridden; 
    \item Inheritance: the inheritance relationship between the classes that the candidate and the Signaler belong to.
\end{itemize}

To extract key elements for this EP, we first collect the unconditional exception information from previously obtained ETS and record its class as $\mathtt{Super}$. 
The class that the candidate belongs to is called $\mathtt{Sub}$.
Then we perform a class hierarchy analysis to extract the path from $\mathtt{Sub}$ to $\mathtt{Super}$ and record it as \textit{Inheritance}.
}

\journal{
\textbf{EP 2: Transfering wrong value to \textit{keyVar}.} 
This EP denotes that the given candidate transfers a wrong value to the $\mathit{keyVar}$ in \textit{crashAPI}, whose value can influence the triggering of the target exception.
As shown in Listing~\ref{fig: EP2_1}, the application method $\mathtt{crashMethod()}$ passed the value of $\mathtt{y}$ to $\mathtt{crashAPI()}$ as a parameter, which is a $\mathit{keyVar}$. 
Unfortunately, the passed value is unexpected.
As the value of $\mathtt{y}$ is related to the parameter variable $\mathtt{p2}$, its actual source is method $\mathtt{caller()}$, which transfers the value of $\mathtt{x}$ to the second parameter of $\mathtt{crashMethod()}$, i.e., $\mathtt{p2}$.
}

\begin{lstlisting}[caption={\journal{Transfering Wrong Value to $\mathit{keyVar}$}}, label={fig: EP2_1}]
public class A2{
    public void caller() {  // candidate method
        int x = computeX();
        crashMethod(0, x); // pass wrong value to p2
    }
    public void crashMethod(int p1, int p2){
        int y = p2-1;
        crashAPI(y); // y is the keyVar
    }
}

public class F2{
    public void crashAPI(int crashParameter, long otherParameter) {
        signaler(3, crashParameter);
    }
    public void signaler(int uncheckedParameter, int checkedParameter){
        if(checkedParameter==-1) 
            throw new SomeException();  
    }
}
\end{lstlisting}

\journal{
The key elements to be collected for this EP are as follows: 
\begin{itemize}[leftmargin=10pt]
    \item Signaler: the signature of the last invoked framework method in the crash stack, which throws the target exception; 
    \item CrashAPI: the signature of the first invoked framework API in the crash stack;
    \item KeyVariable: the variable in CrashAPI that can affect the triggering of the target exception;
    \item CallChain$\mathit{_{Crash}}$: the method calling trace from the candidate to the Signaler with the passing information of KeyVariables.
    When using this chain, the start and end points can be customized according to users' requirements.
\end{itemize}

For this EP, we first record the $\mathit{keyVars}$ in the target ETS as KeyVariables.
Then, we collect the method invocation chain from the given candidate to the Signaler as CallChain$\mathit{_{Crash}}$.
and extract the variable-passing chain from Signaler to the CrashAPI as CallChain's additional information.


}

\journal{
\textbf{EP 3: Modifying class fields or objects that can influence \textit{keyVar}'s value.} 
Besides directly transferring a wrong value to $\mathit{keyVar}$, there are other ways that can lead to $\mathit{keyVar}$-related exceptions.
EP~3 denotes that the given candidate can also indirectly influence the value of $\mathit{keyVar}$ by modifying the value of a class field or an object.
For the class fields, as shown in Listing~\ref{fig: EP2_2}, method $\mathtt{crashMethod()}$ invokes $\mathtt{crashAPI()}$, which passes an unexpected value $\mathtt{x}$ related to the class field $\mathtt{f}$.
The value of $\mathtt{f}$ is modified in method $\mathtt{fieldOpMethod()}$, which further causes an unexpected crash.
For the objects, as shown in Listing~\ref{fig: EP2_3}, method $\mathtt{CrashMethod()}$ invoked $\mathtt{CrashAPI()}$ and passed an object $\mathtt{o}$ to it, which is the $\mathit{keyVar}$.
Before the invocation of $\mathtt{CrashAPI()}$, the method $\mathtt{callee()}$ takes the object $\mathtt{o}$ as input and modifies its value, 
which makes the value transferred to $\mathit{keyVar}$ wrong.}

\begin{lstlisting}[caption={\journal{Modifing Class Fields To Influence $\mathit{keyVar}$}}, label={fig: EP2_2}]
public class A3{
    int f = 0; // f is a class field
    public void fieldOpMethod(){ // candidate method
        f = -1; //influence keyVar by modify f
    }
    public void crashMethod(){
        int x = f * 2;
        crashAPI(x); // x is the keyVar
    } 
}
\end{lstlisting}

\begin{lstlisting}[caption={\journal{Modifing Objects To Influence $\mathit{keyVar}$}}, label={fig: EP2_3}]
public class A3_2 {
    public void crashMethod(Obj o) {
        callee(o);
        crashAPI(o);
    }
    public void callee(Obj o) {  // candidate method
        doSomething(o); 
    }
}

public class F3_2 {
    public crashAPI(Obj o) {
        if (SomeCond(o)) 
            throw new SomeException();
    }
}
\end{lstlisting}

\journal{
The key elements to be collected for this EP are as follows: 
\begin{itemize}[leftmargin=10pt]
    \item Signaler: the signature of the last invoked framework method in the crash stack, which throws the target exception; 
    \item CrashAPI: the signature of the first invoked framework API;
    \item KeyVariable: the variable in CrashAPI that can affect the triggering of the target exception;
    \item ModifiedField/ModifiedObject: the name and type of the class field or object variable that is modified by the candidate and affects KeyVariable's value;
    \item EntryAPI: the signature of the method in the stack trace, which uses the ModifiedField/ModifiedObject;
    \item CallChain$\mathit{_{Crash}}$: the method calling trace from the EntryAPI to the Signaler with the passing information of KeyVariables.
    When using this chain, the start and end points can be customized according to users' requirements.
\end{itemize}

For this EP, we first record the $\mathit{keyVars}$ in the target ETS as KeyVariables.
Then, we identify the $\mathit{keyVar}$-related class field and object variables modified by the candidate and label the methods that use them as EntryAPI.
Further, the trace from EntryAPI to CrashAPI is recorded as the CallChain$\mathit{_{Crash}}$, and the variable-passing chain from Signaler to the CrashAPI is extracted as CallChain's additional information.
}

\journal{
\textbf{EP 4: Wrongly invoking specific \textit{keyAPI}.} 
This EP denotes that the given candidate wrongly invokes a framework method, which causes the condition checks before an exception to be satisfied unexpectedly.
As shown in Listing~\ref{fig: EP3}, method $\mathtt{crashMethod()}$ invokes the method $\mathtt{crashAPI()}$, which triggers an exception when the value of class field $\mathtt{f}$ is less than 0. As the method $\mathtt{keyAPI()}$ modifies the value of $\mathtt{f}$ to $\mathtt{-1}$ and method $\mathtt{caller()}$ finally invokes $\mathtt{keyAPI()}$, the method $\mathtt{caller()}$ has responsibility for the unexcepted value of $\mathtt{f}$.}

\begin{lstlisting}[caption={\journal{Wrongly Invoking specific \textit{keyAPI}}}, label={fig: EP3}]
public class A4{
    public void caller(){ // candidate method
        caller2(); 
    }   
    public void caller2(){ 
        keyAPI(); 
    }  
    public void crashMethod(){ 
        crashAPI(); 
    } 
}

public class F4{
    int f = 0;
    public void keyAPI(){ f = -1;} 
    public void crashAPI(){
        if(f<0)  
            throw new SomeException();  
    } 
}
\end{lstlisting}

\journal{
The key elements to be collected for this EP are as follows: 
\begin{itemize}[leftmargin=10pt]
    \item KeyField: the name and type of the class field whose value can influence the exception triggering;
    \item KeyAPI: the signature of the framework API that modifies the value of the KeyField;
    \item CrashAPI: the signature of the first invoked framework API;
    \item CallChain$\mathit{_{KeyAPI}}$: the method calling trace from the candidate to the KeyAPI; 

\end{itemize}

For this EP, we first record the $\mathit{keyAPIs}$ in the target ETS as KeyAPIs, and the $\mathit{keyAPI}$-related $\mathit{keyFields}$ as KeyFields.
Then, we collect the call trace from the buggy candidate to KeyAPI as CallChain$\mathit{_{KeyAPI}}$ by static analysis.
}




\journal{
\subsubsection{\textbf{Collecting Framework Context}}
In the static analysis phase, we have constructed ETSs to represent the framework information.
However, while the statically generated information is useful for code analyzers, it is not easily understandable by humans.
Meanwhile, the semantic-related code information can not be extracted.
To improve the readability of explanations, we make use of LLM to generate explanation-oriented constraints as the framework context.
Given a crash stack trace, for the \textit{keyVar-related} exceptions, we ask LLM to extract the exception-triggering conditions starting from the $\mathit{signaler}$ method, and then transform the conditions into data constraints for the parameters in \textit{crashAPI} step-by-step. 
For the \textit{keyAPI-related} exceptions, besides the data constraints on \textit{crashAPI}, we also ask LLM to summarize the effects of the $\mathit{keyAPI}$ on the \textit{keyField}.
However, LLM may generate hallucinations, i.e., give constraints on a non-existent variable or a bug-irrelevant parameter.
To cope with it, we use the previously generated static analysis results to filter the LLM-generated invalid constraints with three verifiers.
\begin{itemize}[leftmargin=10pt]
    \item \textit{Format verifier}: it verifies whether the variables used in the constraints meet one of the following formats: 1) \textlangle\textit{Parameter \{parameter index\}}: \textit{\{parameter type\} \{parameter name\}}\textrangle; 2) \textlangle\textit{Field \{class name\}}: \textit{\{field type\} \{field name\}}\textrangle.
    \item \textit{SourceCode verifier}: it verifies whether the variables used in the constraints are consistent with the source code, e.g., have the same variable type and name.
    \item \textit{StaticAnalysis verifier}: it verifies whether the parameter variables used in the constraints are consistent with the statically collected variables along the CallChain$\mathit{_{Crash}}$[CrashAPI $\xrightarrow{}$ Signaler], and whether the class fields are in the statically collected KeyField set.
\end{itemize}

For example, Listing~\ref{fig: EP2_1} displays a $\mathit{keyVar}$-related exception and its related methods.
The $\mathit{keyVar}$ is the 1st parameter of $\mathtt{crashAPI()}$. 
By analyzing this code snippet, the static analysis module can give such a variable tracking chain including all the \textit{keyVar-related variables}:
the local variable $\mathtt{x}$ in method $\mathtt{caller()}$ $\xrightarrow{call}$ the 2nd parameter $\mathtt{p2}$ of $\mathtt{crashMethod()}$ $\xrightarrow{call}$ the 1st parameter $\mathtt{crashParameter}$ of $\mathtt{crashAPI()}$ $\xrightarrow{call}$ the 2nd parameter $\mathtt{checkedParameter}$ of $\mathtt{signaler()}$.
The expected framework context is \textlangle Parameter 0: int crashParameter\textrangle $\neq$ -1.
If the variables in the LLM extracted constraints are in the wrong format, e.g. \textlangle\underline{int: }Parameter 0 crashParameter\textrangle $\neq$ -1;
have an incorrect name or type, e.g. \textlangle Parameter 0: \underline{long crashPara}\textrangle $\neq$ -1;
or are not related to any variable in that passing chain, e.g., \textlangle Parameter \underline{1}: long \underline{otherParameter}\textrangle  $\neq$ -1, 
the generated constraint is not qualified.

If the LLM-generated constraints can pass all three verifiers, we take the generated results as a valid framework context, or else, we request LLM to generate constraints in multiple turns with the threshold limit.
A special case is that all the generated constraints fail to pass the \textit{StaticAnalysis verifier}.
Considering that the static analyzer may lose track of parts of the variables that LLM may identify, we try to pick the most representative constraints from all as our target context.
First, we extract the variables involved by a constraint \textit{C} as its variable set $\mathit{C.vars}$.
The partial order relation \textit{R} on \textit{C} is $\preceq$. 
If $C_i.\mathit{vars} \subseteq \mathit{C_j.vars}$, we can say $C_i \preceq C_j$.
For all the generated constraints, if we can find a constraint $C_t$ that $C_i \preceq \mathit{C_t}$ for any $i \neq t$, $C_t$ is the representative constraint that will be used.
If there is no such upper bound, we do not take any constraint as the final framework context.

\subsection{Generating Fault Explanation }~\label{LLMExplain}
Through previous static analysis and the LLM interaction, we have pre-obtained all the necessary code snippets and information, based on which we will further discuss how to generate fine-grained fault explanation reports.

\subsubsection{\textbf{Determining Candidate Explanation Order}}
For LLM, the explanation information for one buggy candidate may be reused to explain another one.
To enhance the clarity of the explanation logic, we arrange the explanation order of the buggy candidates based on their code invocation relationships.
For the $\mathit{keyVar}$-related crash, the order depends on the possible execution trace. 
For methods that exist in the stack trace, the latter to be executed, the earlier to be explained. For other methods not in the trace, they should be explained later than their directly related stack trace method, so we explain them after all the methods in the stack trace have been explained.
For the $\mathit{keyAPI}$-related crash, the order also depends on the possible execution trace.
The latter to call the $\mathit{keyAPI}$, the earlier to be explained. Thus, we also explain them in a bottom-up order.
If a crash is both $\mathit{keyVar}$ and $\mathit{keyAPI}$-related, we explain the $\mathit{keyVar}$-related buggy candidates first, as they provide more environment information when the crash happens.
Besides, for crashes caused by unconditional exception, the closer the inheritance relationship to the class contains the unconditional exception, the earlier to be explained. 

\subsubsection{\textbf{Dynamically Constructing LLM Prompt}}
With a given candidate order, we use LLM to generate the final explanation reports.
The system prompt asks LLM to personate an Android expert to complete explanation tasks using corresponding information, which is listed below.

\prompt{\journalColor{}
\textbf{System Prompt:}
\begin{itemize} [leftmargin=10pt] 
\item You are an Android expert who assists with explaining the crash of the Android application. 
Next, we will provide you with both the crash- and candidate-related information.
According to that as well as the Android basic knowledge, your task is to give a fault explanation to help the developers figure out how the given methods caused the crash.
\item For each crash, we will provide you with the following information: 
\begin{enumerate} 
    \item a crash report that includes the exception type, message, and crash stack trace; {\color{dkgreen}//\textit{Crash Report}}
    \item the exception-triggering constraints on specific Android API. {\color{dkgreen}//\textit{Framework Context}}
\end{enumerate}  
\item For each buggy candidate method., we will provide you with the following information:
\begin{enumerate}  
    \item the code snippet of a crash-triggering candidate method that is detected by a static analysis tool; {\color{dkgreen}//\textit{Code Context}}  
    \item the information about why the static analyzer identifies the candidate method. {\color{dkgreen}//\textit{Candidate Context}}
\end{enumerate}
\end{itemize}
}

Most of the information required by the system prompt can be extracted from CIS's code, framework, and candidate contexts.
However, the original \textit{keyInfo} in CIS only includes raw information about its EP, which cannot be directly utilized by LLM. 
Thus, we use this information to dynamically construct a candidate reason prompt to help LLM better understand the context.
For different EPs, we used the following prompt templates to generate an EP-aware candidate information prompt.

\prompt{
 \journalColor{}
\textbf{Template for EP1:}

The class \textit{Sub} extends class \textit{Super} but does not override the method \textit{Signaler}, which throws an exception unconditionally.
}

\prompt{
 \journalColor{}
\textbf{Template for EP2:}

From the candidate method, wrong values are passed to \textit{CrashAPI} along the call chain $\mathit{CallChain_{Crash}}$[Candidate $\xrightarrow{}$ CrashAPI], which causes the variable $\mathit{keyVar}$ in \textit{CrashAPI} to be unexpected and finally leads to a crash.
}

\prompt{
 \journalColor{}
\textbf{Template for EP3:}

The field \textit{ModifiedField} (or object \textit{ModifiedObject}) is used in \textit{EntryAPI} and then passed along the call chain $\mathit{CallChain_{Crash}}$[EntryAPI $\xrightarrow{}$ CrashAPI] to the method \textit{CrashAPI}.
The candidate method changes the value of \textit{ModifiedField}/\textit{ModifiedObject} and finally leads to a crash.
}



\prompt{
 \journalColor{}
\textbf{Template for EP4:}

The candidate method finally invoked \textit{KeyAPI} along the call chain $\mathit{CallChain_{KeyAPI}}$, which can affect the value of the field \textit{KeyField} in the Android Framework.
The method \textit{CrashAPI} crashes due to the changing of \textit{KeyField} making the exception-triggering constraint satisfied.
}

\subsubsection{\textbf{Generating Fault Explanation Reports}}
With the constructed prompts, we can get the final report. The report for the motivating example in section~\ref{sec:motivating} is displayed below.

\explanation{\journalColor{}

\noindent \textbf{Exception Type}: java.lang.IllegalStateException

\noindent \textbf{Message}: attempt to re-open an already-closed object: SQLiteProgram: SELECT count(\_id) FROM cg\_caches WHERE reason $\ge$ 1

\noindent \textbf{Stack Trace}: acquireReference, 
simpleQueryForLong,
simpleQueryForLong,
getAllCachesCount,
run

\noindent \textbf{Framework Constraint}: [acquireReference]: \textlangle Field SQLiteClosable: int mReferenceCount\textrangle $\ge$ 0

\noindent \textcolor{gray!30}{\rule{\linewidth}{1pt}}

\noindent \textbf{Global Explanation}:
The crash is caused by the improper handling of SQLite prepared statements and the database in the application. Here is a summary of how the candidate methods contributed to the crash:

1. $\mathtt{clearPreparedStatements}$. This method closes all prepared statements and clears the list of statements. If this method is called while any statements are still needed, it leads to ``IllegalStateException'' when those statements are accessed afterward, as their resources are already released.

2. $\mathtt{closeDb}$. This method closes the database and clears all prepared statements. If this method is called while any database operations are still ongoing, it results in an ``IllegalStateException'' because the database and statements are closed while still in use.

3. $\mathtt{simpleQueryForLong}$. This method executes a query using a prepared statement. If the statement has been closed by $\mathtt{clearPreparedStatements}$ or $\mathtt{closeDb}$ before this query is executed, it leads to an ``IllegalStateException''.

4. $\mathtt{getAllCachesCount}$. This method calls $\mathtt{simpleQueryForLong}$ on a prepared statement to get the count of all caches. If the statement has been closed previously, calling this method results in an ``IllegalStateException''. 

 The crash is ultimately caused by attempting to use closed SQLite objects. Ensuring that prepared statements and the database are not closed prematurely or while still in use will prevent this crash.

\noindent \textcolor{gray!30}{\rule{\linewidth}{1pt}}

\noindent \textbf{Candidate Method 1}: $\mathtt{clearPreparedStmts}$

 \textbf{Method Code}:

\quad public static void clearPreparedStatements() \{

    \quad\quad for (SQLiteStatement statement: statements) \{
    
        \quad\quad\quad statement.close();
        
    \quad\quad\}
    
    \quad\quad statements.clear();
    
\quad \}

\textbf{Explanation}:  The clearPreparedStatements method closes all SQLite statements, making them invalid for future use. When the application tries to execute a query using one of these closed statements, it crashes because the statement is already closed.

\noindent  \textbf{Candidate Method 2}: $\cdots$
}

Based on this fault explanation report, users can first get an overview of the crash by reviewing a set of crash-related information, including the exception type, message, stack trace, and framework constraint.
The following part \textit{Global Explanation} mainly introduces how each candidate method contributed to the crash.
It also discusses the application code's behavior when the crash happens, which is produced by integrating information from all the candidates.
Debugging with such information can help users quickly figure out the crashing scenario, which is because the application trying to execute a query using one of these closed statements.
Following that, our report provides detailed information for each candidate, including the candidate's method code and the explanation of why it is likely to be buggy.
For candidate $\mathtt{clearPreparedStmts}$, the report points out it closes all the $\mathtt{SQLite}$ statements, making them invalid for future use.
The crash happens when executing a query using one of the closed statements (in the crashMethod $\mathtt{simpleQueryForLong}$).
}

\section{Evaluation}\label{evaluation1}
We implemented the framework-specific explainable fault localization approach as a prototype \ourtool{}~\cite{CrashTracker}, which consists of about 11 KLoC Java code and 3 KLoC Python code.
It employs the Soot framework~\cite{soot} to analyze Java bytecodes, utilizes Flowdroid~\cite{FlowDroid} to build call graphs, and uses OpenAI's GPT models to generate explanations.
The evaluation of \ourtool{} aims to answer the following research questions.
\begin{itemize}[leftmargin=5pt]
    \item \textbf{RQ1 (Statistic of ETS):} How many ETSs can be extracted from Android framework? What are their characteristics?
    \item \textbf{RQ2 (Effectiveness of \ourtool{}'s fault localization):} Can \ourtool{} effectively identify buggy candidates based on the extraction of ETSs?
    \journal{
    \item \textbf{RQ3 (Effectiveness of \ourtool{}'s fault explanation):} Can \ourtool{} precisely construct CISs and generate debugging-friendly explanations with the power of LLM? 
    \item \textbf{RQ4 (Efficiency of \ourtool{}):} What is the efficiency of the tool \ourtool{}? }
    
\end{itemize}

\subsection{Experimental Setup} \label{sec: setup} 

To answer RQ1, we collected all the major versions of the Android frameworks to extract ETSs from them.  
As the \texttt{android.jar} files in the Android SDK only contain stub methods but not real code implementation, 
we load the published Android images and pull the complete jar files from the system.
In total, we gathered ten versions of framework files for Android, ranging from 2.3 to 12.0~\cite{AndroidFrameworkImpl}. Android 3 was not included due to its unavailability.

To answer RQ2, the Android framework-specific crash dataset is required.
There are two off-the-shelf benchmarks related to Android crash datasets.
The first one is the dataset extracted by Fan et al.~\cite{DBLP:conf/kbse/FanSCMLXP18}, which includes 194 crashes from GitHub~\cite{Github} issues. 
Another one is ReCBench~\cite{DBLP:conf/issta/KongLGBK19}, which contains more than 1,000 crashes from real-world app execution.
To focus on the framework-specific crashes only, Kong et al.~\cite{DBLP:journals/ase/KongLGRZBK21} filtered these two datasets with the following criteria: 1) the stack trace must contain at least one application-level method; 2) the \textit{signaler} must be a framework-level method.
After filtering, it extracted 500 crashes (D500) on ReCBench~\cite{DBLP:conf/issta/KongLGBK19} and 69 crashes (D69) in Fan et al.'s work~\cite{DBLP:conf/kbse/FanSCMLXP18}, 
and then divided the 569 apps into three buckets according to the location of the \textit{buggyMethod} 
(Category A: \textit{buggyMethod} in $T_{crash}$;
Category B: \textit{buggyMethod} in $T_{execute}$ but not in $T_{crash}$; 
Category C: crash arises from non-code reasons).

We reuse and update these 569 crashes as our evaluation dataset by 1) adding oracle information about both the \textit{keyCondVars} and \textit{externalVars}; 2) pointing out the code-level \textit{buggyMethods} for crashes in category C and recording their original labels as extra non-code characteristics.
In addition to the 569 collected Android APKs, we also take the Android third-party SDKs as the application-level code.
We search the SDK library projects on GitHub that have large star numbers, active commit behavior, and normalized issue submission specifications.
By manually reviewing, we pick two popular projects, facebook-android-sdk~\cite{facebook} and google-map~\cite{google-map}.
Then, we filter issues with the keyword \textit{is:issue is: Closed ``AndroidRuntime'' OR ``Crash''}.
For the 84 + 53 issues matched, we manually check whether the crash is Android framework-specific and whether a fixing commit is given.
Overall, 11 framework-specific crash reports with fixing revision (D11) are collected.
They form a large dataset with 580 crash reports (D580).

For SOTA tool comparison, as the CG-tracing-based Java fault localization tool CrashLocator~\cite{DBLP:conf/issta/WuZCK14} is not available, we implement a similar strategy (b1-ExtendCG) in our tool and perform self-comparison.
Besides, \textit{Anchor}~\cite{DBLP:journals/ase/KongLGRZBK21} is a novel Android framework-specific fault localization tool, which first applies machine learning algorithms to categorize each new crash into a specific category (A/B/C), and then combines the application-level static analysis and similar crash query to achieve the final buggy ranking. We also compare with $Anchor$ using the 569 test cases.

\journal{

RQ3 aims to evaluate the effectiveness of the explanations generated by \ourtool{}, which integrates the static analysis and LLM.
First, we focus on the precision of the constructed CISs.
1) For the \textit{candidate context}, we evaluate the correctness of keyInfo extracted by static analysis.
To perform that evaluation, we collect all the candidate explanation reports and group them according to their explanation patterns. For each type, 20 cases are randomly picked via uniform sampling. 
If there are fewer than 20 instances for a specific type, all the instances within that type will be selected. 
In total, we get 66 cases for candidate context evaluation. 
2) For the \textit{framework context}, we evaluate the correctness of constraints extracted by LLM.
On the one hand, we evaluate the correctness of all the successfully generated framework contexts, whose number is 225, with the latest LLM model \textit{GPT-4-1106-preview}, which represents the most advanced capabilities of LLM (according to the chatbot arena leaderboard's data\cite{chiang2024chatbot}, as of March 1, 2024). 
On the other hand, we compare the correctness of the constraints extracted upon both plain LLM and our approach using a classical LLM model \textit{GPT-3.5-turbo}, which is more cost-effective.
For the plain-LLM approach, only the crash report and the stack-trace-related code snippets are provided, while for our approach, three types of contexts are provided to LLM as CISs.

We also conduct a questionnaire survey to assess developers' preferences regarding both the fine-grained explanations provided in \ourtool{}'s final report and the naive report produced solely by \ourtool{}'s fault localization module. 
To enhance the diversity of the reports in the survey, we grouped reports with the same first five layers of the crash stack and selected up to three reports from each group. This resulted in 44 groups of explanation reports, each of which contains a final report and a naive report. Each report was evaluated by two developers, leading to a total of 44*2*2 = 176 cases. Eight developers, each with over three years of Java experience (four of whom have extensive Android development experience), reviewed all the cases.
Before conducting the survey, we held a brief training session to introduce the necessary knowledge about the survey, including the debugging process with an explainable fault localization report, the metrics to be scored, and the directory structure of the distributed folder. Additionally, a README file was provided to give a brief illustration. After the meeting, each survey participant independently completed the questionnaire.

Finally, in RQ4, we evaluate the efficiency of \ourtool{}'s two key modules: fault localization and explanation.
Our analyses are performed on a Linux server with two Intel® Xeon® E5-2680 v4 CPUs and 256 GB of memory.
For LLM, we use the \textit{gpt-4-1106-preview} model to execute by calling OpenAI's APIs.
}

\subsection{RQ1: ETS Construction on Android Frameworks} \label{sec: ExceptionExtraction}
\begin{figure}[!htb]
    \setlength{\abovecaptionskip}{5pt}
    \setlength{\belowcaptionskip}{-10pt}
    \centering
    \includegraphics[width=0.48\textwidth]{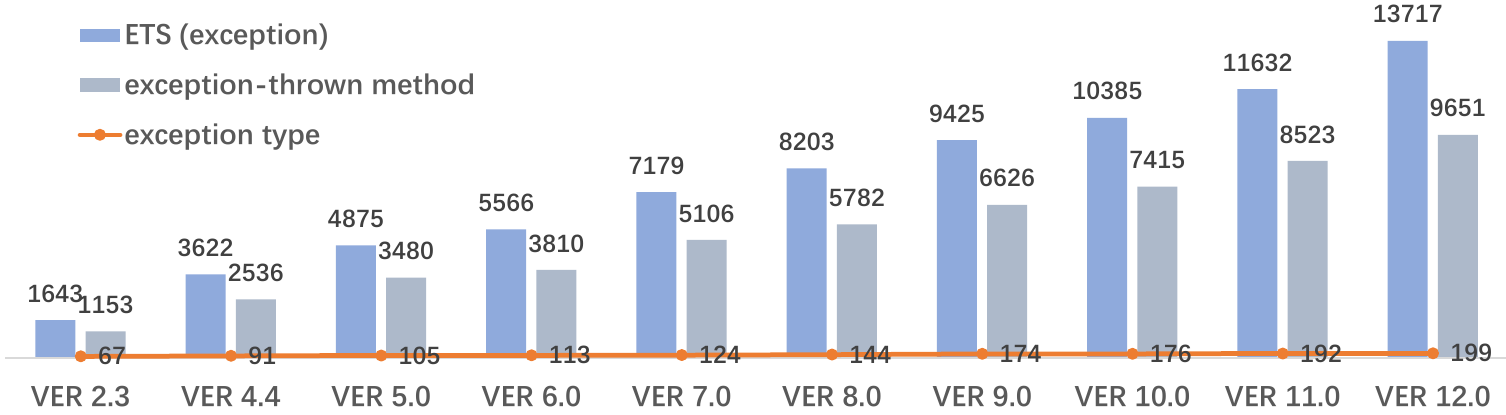}
    \caption{ETSs in Multiple Android Frameworks}
    \label{fig: CrashTracker-excptionNumber}
\end{figure}

The Android framework updates rapidly, so the exceptions may also change with time.
Fig.~\ref{fig: CrashTracker-excptionNumber} shows the number of ETS in different versions of the Android framework, in which each ETS denotes one unique exception.
With the evolution of the framework, the number of exceptions, exception types, and exception-thrown methods all increase.
For example, in Android 10.0, 10,385 exception-thrown methods throw 7,415 exceptions with 176 types.
Among them, 6,917 ETSs have non-empty message descriptions, of which 29.0\% of them contain non-constant values in the generated regular expression.

For each exception, we trace its \textit{keyConds}, \textit{keyCondVars}, and the \textit{externalVars}.
When tracing all the key conditions related to an exception, the average condition length is 2.78.
So we set the threshold in \textit{keyCond} collection as 3.
Table~\ref{tab: RelatedCondition} shows the distribution of the ETSs with different ETS-related types.
We give the statistical results on the oldest version (Android 2.3), the latest version (Android 12.0), and the average value for all ten versions.
According to the results, most ETSs only have \textit{keyVars} (42\%) or \textit{keyAPIs} (22\%). 
Around 18\% ETSs have both \textit{keyVars} and \textit{keyAPIs} and 4\% ETSs do not have \textit{keyCond}.
About 15\% ETSs could not link with any \textit{externalVars}, most of which are from the caught and re-thrown exceptions or the inter-procedural-call-related conditions.
On average, we can get 7,625 ETSs from the Android framework code.
Among them, there are 1,859 ETSs that contain 10,227 \textit{keyVar} records and 2,877 ETSs with 308,852 \textit{keyAPIs}.
For the \textit{keyAPIs}, 81,872 are declared in the same class of \textit{signaler}, and 226,980 are located in different classes.

\begin{table}[htp] 
    \setlength{\abovecaptionskip}{5pt}
    \setlength{\belowcaptionskip}{-5pt}
	\caption{Statistic of Each ETS-related Type}\label{tab: RelatedCondition}
	\begin{tabular}{c|c|c|c|c|c|c}  
		\hline \hline 
        \rowcolor{mygray}
        \textbf{\tabincell{c}{Ver}}&  \textbf{\tabincell{c}{\#ETS}}&
         \textbf{\tabincell{c}{key-\\Var}} & \textbf{\tabincell{c}{key-\\API}} & \textbf{\tabincell{c}{keyVar\&\\keyAPI}}
           & \textbf{\tabincell{c}{No key-\\CondVar}} &\textbf{ \tabincell{c}{No exter-\\nalVar}}\\ \hline 
        2.3  	&1,643  &668	 &299	&217	&37	    &122        \\ \hline
        12.0   	&13,717 &5,129	 &2,527	&1,022	&326	&619        \\ \hline
        Avg    &7,625  &2,946 	 &1,245 &705 	&170    &356  \\ \hline\hline
	\end{tabular} 	
\end{table}

Before evaluating the final crash localization results, we first check the correctness of the analyzed ETS-related type.
To achieve this, we review the exception-triggering code snippets related to the collected 580 crash reports.
Two experienced Java developers read the crash trace information and retrieved the corresponding exception in the source code of the Android framework. By manual analysis, they record the key conditions, the key condition variables, and the external variables for all the exceptions triggered in D580.
By comparison, we find that \ourtool{} can correctly identify the ETS-related types (refer to Table~\ref{tab: Strategies}) for 95.5\% crashes. 
There are 26 ones that are misidentified, of which 12 are exceptions in the Android support libraries, which are not collected as framework code; 
three crash reports provide empty message information, which makes the message matching fail. 
Three involve inter-procedural tracing.
The others are condition-related, including one lack of conditions due to the length limit, three getting unrelated \textit{not-return} conditions, two re-throwing a catch condition with unknown \textit{externalVars}, and two wrongly taking synchronized variable or the final constant field as \textit{externalVars}, which actually will not influence the exception-triggering.

\begin{figure}[!htbp]
    \setlength{\abovecaptionskip}{5pt}
    \setlength{\belowcaptionskip}{-15pt}
    \centering
    \includegraphics[width=0.48\textwidth]{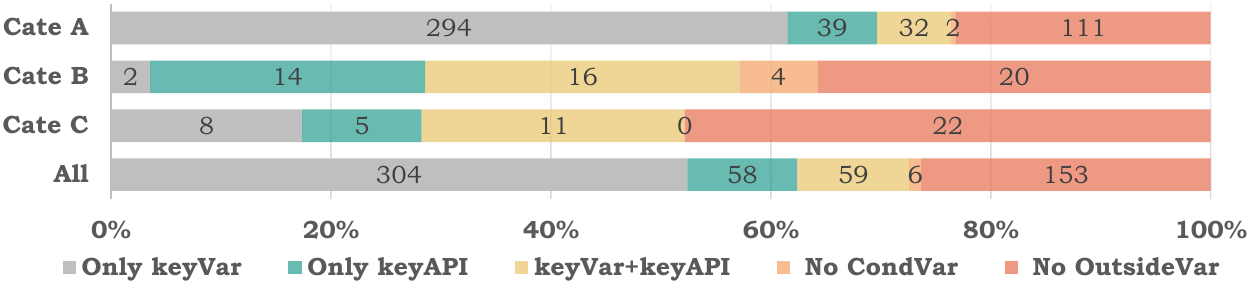}
    \caption{ETS-related Types on Each Category }
    \label{fig: CrashTracker-varAndCate}
\end{figure}

Moreover, we analyze the relationship between the ETS-related types and the crash categories in D580.
As shown in  Fig.~\ref{fig: CrashTracker-varAndCate}, most crashes in category A only have \textit{keyVars}, which is consistent with the \textit{in-stack} behavior of crashes in category A. 
And for category B, whose buggy methods are \textit{out-of-stack}, there are more crashes that have \textit{keyAPIs} than in other categories. \textbf{This reflects the \textit{keyVar} and \textit{keyAPI} analysis work well on both the in-stack and out-of-stack crashes.}

\begin{table*}[tbp] 
\centering  
\tabcolsep=0.5em
\setlength{\abovecaptionskip}{5pt}
\setlength{\belowcaptionskip}{-5pt}
\caption{Effectiveness of \ourtool{} with Multiple Strategies} \label{tab: Strategies}
\begin{tabular}{l|c|c|c|c|c|c|c|c|c|c|c|c}  
\hline \hline
\rowcolor{mygray}
\multicolumn{1}{c|}{\textbf{Strategy}}
&\multicolumn{3}{c|}{\textbf{Statistic}}&\multicolumn{4}{c|}{\textbf{All (580)}} 
&\multicolumn{4}{c|}{\textbf{CategoryB (56)}} & \multicolumn{1}{c}{\textbf{Relationship}} \\ 
\rowcolor{mygray}
&\#Find    &RankSum& CandiAvg & \#R@1 & \#R@5 & \#R@10   & MRR& \#R@1 & \#R@5 & \#R@10  & MRR  & \textbf{of Strategies}  \\ \hline 
\textbf{\ourtool{}}    &568       &954   & 6.35  &500   &562    &567    &0.91     &14     &38     &43     &0.44 &
\multirow{9}{*}{\begin{minipage}[b]{0.27\columnwidth}
    \centering \raisebox{-0.8\height}{\includegraphics[width=\linewidth]{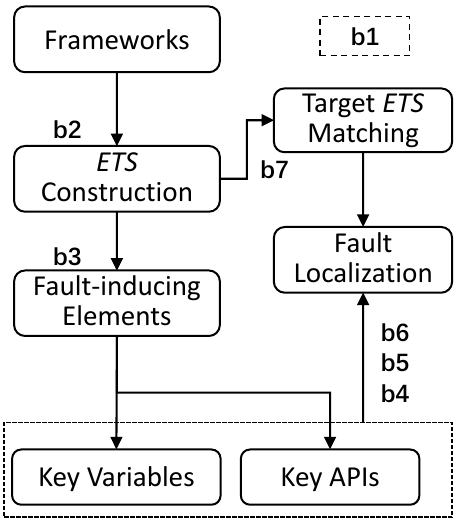}}
\end{minipage}}    \\ \cline{1-12}
b1-ExtendCG   &-8   &+564 &+20.63 &-5    &-11    &-13    &-0.02    &-8     &-7    &-11    &-0.16 & \\ \cline{1-12}
b2-AllConditon      &-0      &-0          &+0.42  &-0    &-0     &-0     &-0.00    &-0     &-0     &-0     &0.00      &     \\ \cline{1-12}
b3-NoCondType &-15  &+409  &-0.21  &-4    &-12    &-15    &-0.02    &-8     &-11    &-14    &-0.19& \\ \cline{1-12}
b4-NoKeyVar         &-3     &+66         &-0.15  &-0    &-3     &-2     &-0.01    &-0     &-3     &-2     &-0.06       &   \\ \cline{1-12}
b5-NoKeyAPI   &-11  &+182 &-1.18  &-4    &-8     &-11    &-0.01    &-4     &-7    &-10     &0.10& \\ \cline{1-12}
b6-NoCallFilter     &-0    &-0           &+0.84  &-0    &-0     &-0     &-0.00    &-0     &-0     &-0     &-0.00       &   \\ \cline{1-12}

b7-Version2.3   &-7   &+219 &-2.57  &-3    &-5     &-8     &-0.01    &-3     &-5     &-8     &-0.10&    \\ \cline{1-12}
b7-Version8.0  &-0     &+4  &-0.01  &-2    &-0     &-0     &-0.001   &-0     &-0     &-0     &-0.00       & \\ \hline\hline
\end{tabular}
\end{table*}

\subsection{RQ2: Effectiveness of \ourtool{}'s Fault Localization} \label{sec:Buggylocalization}
To evaluate the effectiveness of \ourtool{}'s fault localization module, we first made a group of self-comparisons. The results are displayed in Table~\ref{tab: Strategies}.
The second column displays how many \textit{buggyMethod} can be located by the tool in its candidate list. 
The following \textit{RankSum} gives the cumulative sum of the ranking of \textit{buggyMethod} in the candidate list.
If the \textit{buggyMethod} is not found, we use $\mathit{max(candiSize+1, n)}$ as its ranking value, in which we suppose users have to look up at least $n$ candidates to find the target method ($n=20$ by default).
The column \textit{CandiAvg} gives the number of provided candidates on average.
The following eight columns give the precision evaluation results on D580, especially on the crashes with category B.
Two metrics, \textit{Recall} and \textit{Mean Reciprocal Rank (MRR)}~\cite{MRR} are used, in which \#R@N counts the number of reports that rank \textit{buggyMethod} in its first N candidates and MRR denotes the mean of the multiplicative inverse of the rank of the first correct location. It can be calculated by the formula $MRR = \frac{1}{E} \sum_{n = 1}^{E}\frac{1}{Rank_i}$.

 The first line in Table~\ref{tab: Strategies} gives the default results of \ourtool{}, which can find correct \textit{buggyMethod} for 568/580 crashes.
 For \ourtool{}, its \textit{RankSum} is 954. 
 \ourtool{} can provide a compact list with only 6.35 candidates.
 For all the 580 crashes, \ourtool{} has high precision at R@1, R@5, and R@10.
 For crashes in category B, it can still find out most of the \textit{buggyMethod} with a few candidates, e.g., 68\% for 5 candidates and 77\% for 10 candidates.
 Besides the code-level localization, we also label the non-code reasons as buggy tags for 45 crashes located in category C.
 Comparing the given labels, our non-code reason analyzer can correctly infer 27/45 already labeled tags, and we can observe another 40 tags that are not labeled in the original dataset.

\begin{table}[!bp] 
	\centering  
    \tabcolsep=0.45em
    \setlength{\abovecaptionskip}{5pt}
    \setlength{\belowcaptionskip}{-5pt}
    \caption{Precision on Different ETS-related Types} \label{tab: Effectiveness2}
	\begin{tabular}{l|c|c|c|c|c}  
		\hline \hline 
        \rowcolor{mygray}
        \textbf{Source Type}   &\textbf{Count}  & \textbf{\tabincell{c}{R@1(\%)}} & \textbf{\tabincell{c}{R@5(\%)}} & \textbf{\tabincell{c}{R@10(\%)}} & \textbf{MRR} \\ \hline 
    No keyCondVar &6     &0.83   &1.00   &1.00   &0.92 \\ \hline
    No externalVars   &153   &0.77   &0.92   &0.94  &0.83  \\ \hline
    Only have keyVar      &304   &0.97   &1.00   &1.00   &0.98 \\ \hline
    Only have keyAPI      &58    &0.74   &0.91   &0.96   & 0.82\\ \hline
    \tabincell{l}{Both have keyVar\\ and keyAPI}    &59    &0.68   &0.98   &0.98   &0.80  \\ \hline \hline 
	\end{tabular}
\end{table}

Then, we compare \ourtool{} with a set of variants.
In strategy $b1$, we implement a CG-expansion-based approach~\cite{DBLP:conf/issta/WuZCK14} to trace invoked methods along the call edges with a call depth of 5.
However, this approach generates too many candidates, which increases by 20.6 candidates for a crash on average.
And it can only locate (43-11)/56 = 57\% \textit{buggyMethod} in category B with 10 candidates.
The strategy $b2$ is used to evaluate whether tracking conditions with length three influence the results.
In this strategy, all the conditions will be collected. However, it did not bring higher precision but reported a bit more candidates.
In $b3$, we suppose the ETS information is not available, i.e., only strategy $S_2$ is adopted, which decreases the overall precision, especially for cases in category B.
Moreover, in Table~\ref{tab: Effectiveness2}, we present the precision of \ourtool{} on these ETS-related types, 
including the number of cases under test (count), the ratio of reports that rank \textit{buggyMethod} in the first N candidates ($R@N(\%)=\frac{\#R@N}{count}$), as well as the MRR. 
From the results, only the crashes relating to \textit{keyVars} are easier to locate, as most of them exist in the stack.
For type \textit{No keyCondVar}, we can quickly locate the method that needs to be overridden.
And the \textit{keyAPI}-related crashes and those with unknown \textit{externalVars} are difficult to locate with one chance.
Overall, the precision improves much on R@1(\%) to R@5(\%) which indicates the effectiveness of \ourtool{} with a compact candidate set.
Strategies $b4$ and $b5$ are designed to validate the effectiveness of \textit{keyVar} and \textit{keyAPI} identification, respectively.
All of them decrease the overall precision and increase the sum of the ranking value.
In $b_6$, we do not filter the candidate methods that have too many callers. The results show that the filtering strategy reduces parts of ineffective candidates.
Finally, $b_7$ does not match target ETS from multiple versions and only considers a fixed version. For example, when using the fixed versions 2.3 and 8.0, parts of crashes cannot match their target exception.
According to the results, \textbf{both the ETS construction and the multiple strategies contribute to precise fault localization.}

After the self-comparison, we compare the effectiveness of \ourtool{} with the state-of-the-art tool \textit{Anchor}.
Table~\ref{tab: Effectiveness} gives the R@N(\%) and MRR results of Anchor and \ourtool{} on the datasets D500 and D69. 
Overall, \textbf{\ourtool{} achieves improvement in precision on both datasets}, i.e., achieves 7.4\%, 11.5\%, and 12.6\% improvement over Anchor on R@1, R@5, and R@10 metrics, and improves MRR from 0.84 to 0.91.
One special case is the result of R@1 in category B, which also influences the MRR. 
For these out-of-traces crashes, \ourtool{} fails to find out the \textit{buggyMethod} in the first place compared to Anchor.
But if we review the first five candidates, our tool can actually locate more \textit{buggyMethods}.
The reason is that the crashes in category~B are usually triggered due to the invocations of \textit{keyAPIs} related to the \textit{signaler} method.
The most common usages are pairwise API operations, e.g., \texttt{register()} and \texttt{unregister()}. 
It is difficult to know whether the \texttt{register()} is redundant or the \texttt{unregister()} is missed without understanding the developers' intention.
Though it brings parts of FPs when only the top one candidate is checked, we provide the complete call paths from each candidate to the \textit{signaler} in the bug report to help make quick confirmation among multiple candidates.

\begin{table}[htbp] 
    \centering  
	\setlength{\abovecaptionskip}{5pt}
    \setlength{\belowcaptionskip}{-5pt}
    \tabcolsep=1em
    \caption{Comparison with SOTA Fault Localization Tool} \label{tab: Effectiveness}
	\begin{tabular}{l|c|c|c|c}  
		\hline \hline 
        \rowcolor{mygray}
        \multicolumn{5}{c}{\textbf{Anchor}} \\ \hline
        \rowcolor{mygray}
        \textbf{Dataset}    & \textbf{\tabincell{c}{R@1 (\%)}} & \textbf{\tabincell{c}{R@5 (\%)}} & \textbf{\tabincell{c}{R@10 (\%)}} &\textbf{MRR} \\ \hline 
        D500-A   &0.90   &0.91   &0.91   &0.90   \\ \hline
        D500-B   &0.37   &0.59   &0.61   &0.46   \\ \hline
        D500-C   &0.72   &0.75   &0.75   &0.73   \\ \hline

        D69-A    &0.72   &0.93   &0.93   &0.81   \\ \hline
        D69-B    &0.43   &0.43   &0.43   &0.43   \\ \hline
        D69-C    &0.25   &1.00   &1.00   &0.40   \\ \hline

        D569     &0.81   &0.87   &0.87   &0.84   \\ \hline  \hline 
	\end{tabular}
    \begin{tabular}{l|c|c|c|c}  

        \rowcolor{mygray}
        \multicolumn{5}{c}{\textbf{\ourtool{}}} \\ \hline 
        \rowcolor{mygray}
        \textbf{Dataset}    & \textbf{\tabincell{c}{R@1 (\%)}} & \textbf{\tabincell{c}{R@5 (\%)}} & \textbf{\tabincell{c}{R@10 (\%)}} &\textbf{MRR} \\ \hline 
        D500-A   &0.96           &1.00   &1.00 &0.97    \\ \hline
        D500-B   &\textbf{0.22}  &0.67   &0.78  &\textbf{0.42}   \\ \hline
        D500-C   &0.95  &1.00   &1.00 &0.98   \\ \hline

        D69-A    &0.78  &1.00   &1.00  &0.87   \\ \hline
        D69-B    &0.43  &0.71   &0.71  &0.55   \\ \hline
        D69-C    &0.75  &1.00   &1.00  &0.83   \\ \hline

        D569     &0.87  &0.97   &0.98  &0.91   \\ \hline \hline 
	\end{tabular}
\end{table}


Then we investigate the reasons that lead to imprecision in static-analysis-based fault localization.
The false positive (FP) candidates denote the \textit{non-buggyMethods} that are reported in the candidate list.
As we have a compact buggy candidate set, the average FP number for each case is 5.35 on average. 
There are two key reasons that make them reported.
First, our approach uses static analysis to compute the \textit{keyConds}, based on which we further get the \textit{keyVars} and \textit{keyAPIs}.
The imprecision in the static analysis, e.g., the FPs in CG construction, will lead to misidentified information in ETS and finally influence the fault localization results.
Besides, even though conditions are collected, we do not combine the constraint-solving techniques to reduce the ineffective results, which will be explored in further work.
Second, considering there may be misidentified ETS or ETS whose triggering information is unknown, 
we add the application-level methods in the crash stack as the default candidates conservatively, which also brings some FPs.

The false negative (FN) candidates denote the \textit{buggyMethods} that are not in the candidate list. 
One reason is also the imprecision in the static analysis.
Meanwhile, the exceptions with incomplete ETS information, e.g., caught from try-catch blocks or native methods, may bring FNs. 
Also, we collect at most three conditions for an exception and filter the candidates whose callee has too many callers.
These settings make a balance between efficiency and effectiveness but may cause unexpected FNs.
For the 580 cases in D580, we can find 568 \textit{buggyMethods} and miss 12 ones.
Among them, 8 are related to native signaler methods, one suffers from unknown crash message information, 
and two are missed for lacking implicit data flow relationship, e.g., when the signaler \textit{android.widget}.\textit{Spinner.setAdapter} is invoked, another method must be overridden as its default value will lead to a crash.

\journal{

\begin{table}[!bp]
\journalColor{}
\centering  
\caption{\journalColor{} Correctness of Candidate Context in \textit{CIS}}\label{tab: RQ3.1}
\begin{tabular}{l|l|c|c|c}  
    \hline \hline 
    \textbf{Pattern} & \textbf{Key Element} & \textbf{Correct} & \textbf{Total} & \textbf{Accuracy} \\\hline 
    \multirow{1}{*}{EP 1}  
        & Inheritance & 6 & \multirow{1}{*}{6} & 1.00 \\
    \hline
    \multirow{2}{*}{EP 2} 
        & KeyVar & 19 & \multirow{2}{*}{20} & 0.95 \\    
        & CallChain$\mathit{_{Crash}}$ & 19 & & 0.95 \\    
    \hline
    \multirow{4}{*}{EP 3}
        & KeyVar & 18 & \multirow{4}{*}{20} & 0.90 \\
        & ModifiedReference & 17 & & 0.85 \\
        & EntryAPI & 20 & & 1.00 \\
        & CallChain$\mathit{_{Crash}}$ & 20 & & 1.00 \\
    \hline
    \multirow{4}{*}{EP 4}
        & KeyField & 15 & \multirow{4}{*}{20} & 0.75 \\
        & KeyAPI & 19 & & 0.95 \\
        & CrashAPI & 20 & & 1.00 \\
        & CallChain$\mathit{_{KeyAPI}}$ & 19 & & 0.95 \\
    \hline 
    All & - & 56 & 66 & 0.85 \\
    \hline \hline 
\end{tabular} 
\end{table}

\subsection{RQ3: Effectiveness of \ourtool{}'s Fault Explanation}  \label{sec: RQ3}
This part evaluates the effectiveness of \ourtool{}'s fault explanation module.
First, Table~\ref{tab: RQ3.1} shows the evaluation results of the \textit{keyInfo} in the collected 66 candidate information summaries (CISs).
The first two columns give the explanation pattern (EP) and its key elements. 
Different EPs' key elements vary, so we check the correctness of all the key elements.
However, to save space, the accuracy of the elements that can be directly obtained from the crash report, including the Signaler and the CrashAPI, are not listed.
The third column gives the number of correctly identified information and the following column gives the total number of CIS collected for this pattern.
The final column presents the accuracy rate. 
The mean accuracy rate for EP is 56 / 66 = 0.85, i.e., 56 cases are considered accurate as all of their key elements are correct.
The results show that the static analysis module can provide practical information for CIS construction.
Totally, 10 cases are incorrect, the root causes of which can be categorized into two types.
First, \ourtool{} makes a conservative static analysis when encountering methods that cannot be analyzed, e.g., native code, which produces six error cases.
Secondly, \ourtool{} may lose precision if the given case does not conform to our assumptions. For example, our analysis assumes the crash stack has only two layers of method calls, one framework layer and one application layer.
However, due to Android callbacks, some cases may have multiple layers of interlaced calls in the crash stack, leading to four errors.

Further, we assess the capability of the LLM in extracting constraints, i.e., the framework context. 
We are unable to map parts of methods to a single method code snippet because the signature's parameter information is missing and there are multiple overridden methods, due to limitations in the datasets we collected. 
Besides, some methods's code snippets can not be obtained, as they are native methods that correspond to C or C++ implementations.  
So, we filtered crashes with the above characteristics and obtained 225 cases for the constraint extraction evaluation. 
Table ~\ref{tab: RQ3.2} provides the evaluation results, in which the second and third columns give the number of the correctly and wrongly extracted constraints, and the last column gives the accuracy rate.
The constraint analysis in \ourtool{} is divided into two stages: \textit{constraint extraction} and \textit{constraint propagation}. In the first stage, we provide LLM with the \textit{signaler} method and ask the LLM to extract the exception-triggering-related constraints as the extracted constraint \textit{ec}.
In the second stage, we provide the LLM with both the \textit{ec} and the code snippet of \textit{signaler}'s caller methods. Then we ask the LLM to transform \textit{ec} a new constraint along the framework level call stack as the propagated constraint \textit{pc}.
After these two stages, the final constraint \textit{fc} can be obtained. 
It is worth noting that, due to the strong semantic understanding ability of LLM, sometimes, a wrong constraint occurring in the earlier stages does not necessarily reflect on the correctness of \textit{fc}.
In Table ~\ref{tab: RQ3.2}, the directly extracted constraints are usually correct with an accuracy of 99\%.
Due to the complexity of the constraint propagation, the accuracy for the final constraints is 63\%.
For about 60\% cases, the constraints extracted in each phase are all correct.

\begin{table}[htbp]\journalColor{}
\centering
\caption{\journalColor{} Correctness of Framework Context in CIS}\label{tab: RQ3.2}
\begin{tabular}{l|c|c|c|c}  
    \hline \hline 
    \textbf{}            & \textbf{Correct} & \textbf{Wrong} & \textbf{Total} & \textbf{Accuracy} \\
    \hline
    Extracted Constraint & 224              & 1              & 225            & 0.99                \\
    \hline
    Propagated Constraint  & 140              & 85             & 225          & 0.62                \\
    \hline
    Final Constraint     & 141              & 84             & 225            & 0.63               \\
    \hline
    All Constraint       & 136              & 89             & 225            & 0.60    \\
\hline \hline 
\end{tabular}
\end{table}

We then review the failed cases in the above two phases.
In the constraint extraction phase, one inaccuracy was identified as it missed part of the complete path.
In the constraint propagation phase, there are 85 wrong cases.
This stage requires the LLM to understand the original constraints and make a successful transformation.
The \textit{pc} is labeled as correct only if all the constraints on the propagation chain are correct.
As the length of the average propagation chain is 4.76, the difficulty of this phase increases a lot more.
By investigating the failed constraints, we found that most mistakes are due to incorrect constraint variables and conditions, where incorrect variable is the key reason for imprecision.
There are two constraint variable types: the method parameter and the return variable.
Among the incorrect cases, 24 cases have redundant constraints on parameters that do not have a data dependency relationship with the crash, or lost constraints on parameters that are necessary for crash triggering; and 51 cases give wrong constraints whose return value of a method is incorrect. 
Besides, 10 cases give constraints that have correct variables but wrong constraint checking conditions.


We also compare the correctness of the constraints extracted upon both plain LLM (denoted as \textit{plainLLM}) and our approach (denoted as \textit{\ourtool{}}).
To ensure the evaluated exception-throwing code snippets are distinct, we classify all the cases according to their exception types and exception-throwing statements. 
Only one case is picked for each type. So, we get 34 cases under evaluation from the 225 cases collected before.
To avoid the randomness of LLM, we conducted three rounds of evaluations. 
A case may be discarded in each turn if the extracted framework context fails to pass the verifiers in \ourtool{}.
In total, there are 94 cases in the all-rounds analysis.
Table~\ref{tab: RQ3.3} displays the evaluation results, in which the first three rows give the results for the three rounds, and the last row gives the results for all rounds.
As evident from our findings, our methodology attains an accuracy rate of 83\%, marking a notable enhancement of 38.33\% in comparison to the plain LLM technique.
Based on these results, we further explore the relationship between the final constraint extraction accuracy and the complexity of code context.
Specifically, we define the total number of statements that change control flow from the beginning of the method to the exception throwing point as the \textit{ExP complexity}. 
When using plainLLM, the precision for cases with an \textit{ExP complexity} below and above 10 is only 0.61 and 0.00, respectively. Similarly, when utilizing \ourtool{}, the precision for cases with an \textit{ExP complexity} below and above 10 stands at 0.83 and 0.22.
It can be observed that the accuracy rate is greatly affected by the complexity of the given code context, and our approach can improve the accuracy even in some complex cases, compared to using plainLLM.

\begin{table}[hptb]\journalColor{}
\centering
\caption{\journalColor{}  Correctness of Framework Context in CIS (compare with \textit{plainLLM})}\label{tab: RQ3.3}
\begin{tabular}{l|l|c|c|c|c}  
    \hline \hline 
\textbf{Round} & \textbf{Method} & \textbf{Correct} & \textbf{Wrong}  & \textbf{Total} & \textbf{Accuracy} \\
\hline
\multirow{2}{*}{R1}         & plainLLM & 17 & 14 & \multirow{2}{*}{31}& 0.55  \\
                           & \ourtool{} & 26 & 5  &     & 0.84                     \\
                           \hline
\multirow{2}{*}{R2}         & plainLLM & 21 & 11 & \multirow{2}{*}{32}& 0.66  \\
                           & \ourtool{} & 27 & 5  &     & 0.84                      \\
                           \hline
\multirow{2}{*}{R3}         & plainLLM & 18 & 13 & \multirow{2}{*}{31}& 0.58  \\
                           & \ourtool{} & 25 & 6  &     & 0.81                 \\
                           \hline
\multirow{2}{*}{R-All}      & plainLLM & 56 & 38 & \multirow{2}{*}{94}& 0.60  \\
                           & \ourtool{} & 78 & 16 &     & 0.83  \\
                           \hline \hline 
\end{tabular}
\end{table}

Then, we use a questionnaire survey to evaluate the developers’ preferences on both the naive explanations produced by CrashTracker’s fault localization module only (denoted as $\mathit{Ep_{naive}}$) and the LLM explanations in the final report (denoted as $\mathit{Ep_{final}}$). 
The questionnaire is designed as follows:
\prompt{
 \journalColor{}
\textbf{Questionnaire:}

\begin{itemize}[leftmargin=10pt] \journalColor{}
    \item (\textbf{Reasonability}) On a scale of 1-5, how reasonable do you find the given explanation?
    \item (\textbf{Readability}) On a scale of 1-5, how easily are you able to read and comprehend this explanation?
    \item (\textbf{Conciseness}) On a scale of 1-5, how concise is the given explanation?
    \item (\textbf{Practicality}) On a scale of 1-5, how helpful is this explanation in resolving the crash issue?
\end{itemize}
}

Fig.~\ref{fig: survey_score} shows the survey results on 176 explanation reports with eight violin plots.
Each violin plot denotes the density distribution of the scoring results, which are evaluated with four survey metrics on either $\mathit{Ep_{naive}}$ or $\mathit{Ep_{final}}$ reports.
\textbf{Overall, the $\mathit{Ep_{final}}$ has achieved a significant improvement in user satisfaction, whose average preference score achieves a 67.04\% improvement over $\mathit{Ep_{naive}}$. 
}
Specifically, for the \textit{Reasonability} metric, the average score of $\mathit{Ep_{naive}}$ is only 2.84, while 96.8\% of $\mathit{Ep_{final}}$'s score is not less than 4.
That means, the statically generated results are usually reasonable, but the LLM can still improve the reasonability results by integrating information from multiple dimensions.
For the \textit{Readability} metric, $\mathit{Ep_{naive}}$ mainly scores between 1 and 3, while 98.7\% of $\mathit{Ep_{final}}$'s score is 4 or 5, indicating the readability is obviously improved. 
Regarding the metric of \textit{Conciseness}, the distribution of $\mathit{Ep_{naive}}$ spans from 2 to 5, while $\mathit{Ep_{final}}$ mainly centers around 4. 
This divergence hints at differing preferences among developers when assessing the \textit{Conciseness} metric. 
Despite the shorter explanations in the static analysis results, the LLM-enhanced outcomes receive a slightly higher rating. 
This can be attributed to situations where developers may turn to additional information if they are unable to fully comprehend the information in $\mathit{Ep_{naive}}$. 
Consequently, this overall process might lead them to perceive the results as not truly concise.
For the \textit{Practicality} metric, 96.8\% of $\mathit{Ep_{naive}}$ scores between 1 and 3, while $\mathit{Ep_{final}}$ generally scores between 3 and 5 (96.8\%). 
This is mainly because $\mathit{Ep_{naive}}$ is only generated through templates, making it difficult to combine specific contexts and semantic information to provide effective help for developers.

\begin{figure}[!tb]
    \centering
    \includegraphics[width=0.48\textwidth]{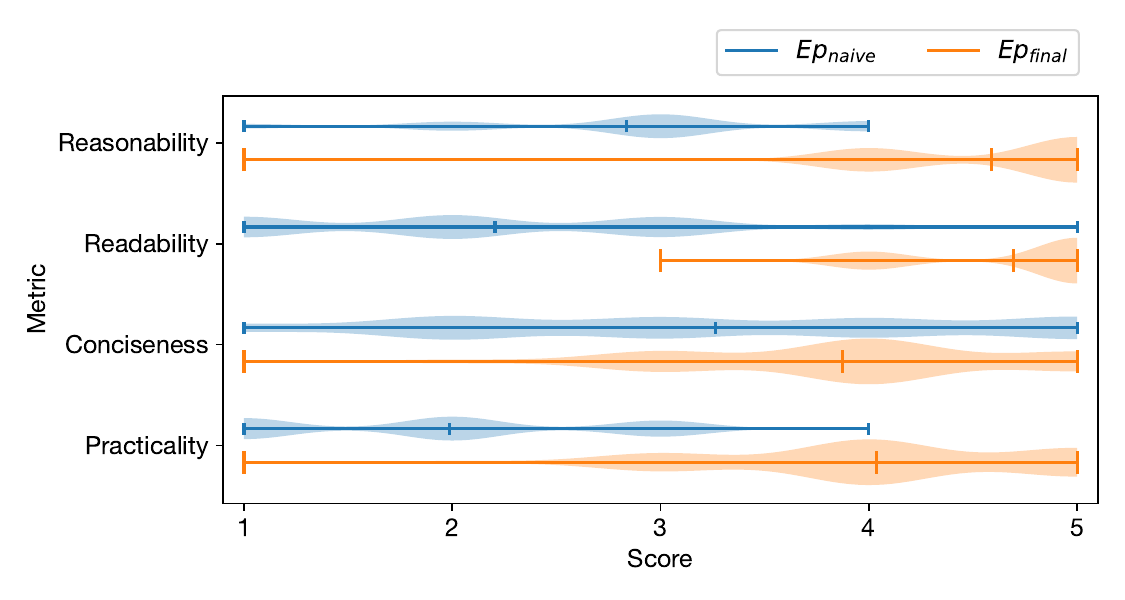}
    \caption{\journalColor{}Survey Results on $\mathit{Ep_{naive}}$ and $\mathit{Ep_{final}}$}
    \label{fig: survey_score}
\end{figure}

\subsection{RQ4: Efficiency} 
In the fault localization phase, \ourtool{} first extracts ETSs for the complete Android frameworks.
It analyzes ten versions of frameworks within 67 minutes, i.e., around 6.7 minutes for one version.
Based on these ETSs, \ourtool{} uses around 95 minutes for 580 apps with 8 threads, i.e., 10 seconds per crash report.
In the fault explanation phase, \ourtool{} first extracts CISs according to static analysis results, which only takes 364 seconds.
Based on these CISs, \ourtool{} used 4.84 hours to generate 225 reports, i.e., 1.29 minutes per crash report. 
Since the ETS construction is a one-time operation, users of our tool will only need to conduct the static fault localization and LLM explanation processes in real time. 
Consequently, the time required for locating and elucidating a single crash report is deemed reasonable (less than 1.5 minutes), which indicates that \textbf{our static analysis and LLM integrated approach can assist developers in quickly identifying buggy methods and obtaining a user-friendly report to aid in their debugging process.}
It is worth noting that a significant portion of the time spent on LLM interactions is dedicated to IO operations. In the current version, our tool waits for the completion of the previous report before initiating the generation of a new one. This step could be parallelized to enhance the efficiency of explanations greatly.
}

\section{Threats to Validity}
The threats to external validity related to the generalizability of the experimental results. 
Although we reuse the two datasets proposed in recent works~\cite{DBLP:journals/ase/KongLGRZBK21,DBLP:conf/kbse/FanSCMLXP18} and add extra crashes related to the third-party-SDKs, the data scale is limited and the crashes are not evenly distributed.
That is because the \textit{keyVar}-related misuses are more common, but the \textit{keyAPI}-related issues are not many, which may be more difficult to find and fix.
As our approach is static-analysis-based, it does not rely on the size of the dataset and achieves a high precision when locating these far-away \textit{buggyMethods}.
We will continuously explore the scalability of \ourtool{} on more crash reports when more datasets are publicly available.
\journal{Meanwhile, limited by the dataset, several \textit{signaler} methods (such as $\mathtt{checkStartActivityResult}$) have a significantly higher proportion, which also affects the diversity of the evaluation on explanation generation. }

Threats to internal validity include the control over extraneous variables.
In our collected datasets, the crash-triggering environment is unknown. 
To maintain randomness, we use the middle version of the matched frameworks, which may bring bias compared with other random strategies.
Several heuristically designed values in candidate ranking are set according to our experience.
Users can adjust them according to their requirements, which does not influence the tool's effectiveness.
Besides, to construct a practical and compact candidate set, we set threshold values during condition collection and caller filtering, which have a weak effect on the results, as evaluated in Table~\ref{tab: Strategies} (b2 and b6).
\journal{Moreover, when evaluating the correctness of constraints on complex code, no pre-defined oracle exists.
When an LLM extracts constraints, it often uses semantic information like crash messages to generate constraints that are closer to semantics rather than being formalized. 
These types of constraints are often more concise and intuitive, but they also hide more details, thereby posing challenges in verifying their correctness.
To enhance the reliability of evaluation outcomes, two authors engage in thorough deliberations for each constrained scenario that fails to yield consistent results during the initial judgment phase.}

\section{Related Work}

\textbf{Crash Trace Based Fault Localization.}
The crash stack trace is the key element in the crash report, based on which, Chen et al.~\cite{DBLP:journals/tse/ChenK15} performed reverse symbolic execution and generated unit test cases.
More works~\cite{DBLP:journals/jss/GuXZZFXQ19,DBLP:conf/issta/WuZCK14,DBLP:journals/ase/KongLGRZBK21,DBLP:conf/icsm/WongXZHZM14,DBLP:conf/icse/ZhouZL12,DBLP:conf/icse/TanDGR18} used crash stack information to narrow down or locate the \textit{buggyMethod}.
The key challenge is that the stack only contains partially executed methods and may not include the buggy one.
So Gu et al.~\cite{DBLP:journals/jss/GuXZZFXQ19} provided an automatic approach to predict whether a crashing-fault resides in a stack trace or not, which denotes the existence of the out-of-stack \textit{buggyMethods}.
To locate them, CrashLocator~\cite{DBLP:conf/issta/WuZCK14} tries to recover the complete execution trace by CG extension on Java projects.
However, without code separation and summary construction, CrashLocator may suffer from a large candidate set or low precision when handling framework-specific crashes.
Compared with it, CrashTracker weakly relies on the CG but relies more on the extraction of \textit{keyVars} and \textit{keyAPIs}. 
To perform stack-trace-based fault-localization on Java programs, both Sinha et al.~\cite{DBLP:conf/issta/SinhaSGJKH09} and Ginelli et al.~\cite{DBLP:conf/qrs/GinelliRMM21} focus on the semantic of exceptions. However, they do not analyze the real exception-thrown points in the frameworks and require manual modeling of specific exceptions, which limits the scalability. For this work, we first perform an automatic semantic analysis for all kinds of framework-level exceptions without any manual modeling. The extracted summaries can help the application-level analysis be more targeted.

To address the problem of Android framework-specific fault localization, researchers combine the learning-based approach with the stack-based analysis.
By learning from similar faults, ExLocator~\cite{DBLP:conf/icse/FanSCMLXPS18} first classifies a crash trace within given exception types and then generates the root causes by static analysis on target applications. 
This tool is not publicly available and only focuses on the given five exception types.
To support more types, Kong et al.~\cite{DBLP:journals/ase/KongLGRZBK21} collected a general dataset with 500 crashing reports for model training and testing.
For a crash trace, their approach first predicts whether the \textit{buggyMethod} exists on the crash stack, then sorts candidates according to the previous classification result.
However, it relies on the labeling of a large-scale dataset, which can not completely cover the numerous and quickly evolving Android framework exceptions.
Compared to it, our tool does not need any prior knowledge of crash fixing and works well for newly detected exceptions.

\textbf{Summary-based Analysis.}
Considering the large size of the framework code, a set of works focuses on how to make an analysis based on the pre-computed summaries.
Some works~\cite{DBLP:conf/icse/PerezL17a,DBLP:conf/ndss/CaoFBEKVC15} noticed that the large-scale framework hinders the inner call relations and hinders the program understanding and debugging.
Among them, Cao et al.~\cite{DBLP:conf/ndss/CaoFBEKVC15} detected the implicit control flow transitions through the Android framework.
Besides, Perez et al.~\cite{DBLP:conf/icse/PerezL17a} generated \textit{predicate callback summaries} for the Android framework and sequence the callbacks.
Recently, Samhi et al.~\cite{DBLP:conf/icse/SamhiBBK21} tried to find out all the atypical methods around ICC in the Android framework. 
To achieve this goal, they retrieve the framework code with a lightweight static analysis.
Though these works are framework-summary-related, none of them focus on exceptions of the Android framework and their summary could not be used as an effective specification to represent the exception-triggering information.

\journal{
\textbf{LLM-based Fault Explanation.}
In recent years, large language models (LLMs) have become increasingly popular in software engineering~\cite{DBLP:journals/tse/WangHCLWW24,DBLP:journals/corr/abs-2308-10620,DBLP:conf/fose-ws/FanGHLSYZ23} owing to their adeptness in producing natural and comprehensive text. 
For the area of explanation, researchers have performed analysis on many aspects, including code understanding~\cite{DBLP:conf/icse/NamMHVM24}, bug explanation~\cite{DBLP:conf/icse/KangYY23}.
Recently, researchers also concentrated on the explanation of fault localization results~\cite{DBLP:conf/wcre/WidyasariANS024,DBLP:journals/corr/abs-2308-15276}, etc.
In 2023, Wu et al.~\cite{DBLP:journals/corr/abs-2308-15276} presented a buggy code and asked LLMs to directly locate the buggy point and explain it.
Instead of designing a new fault localization and explanation approach, this work focuses more on comparing different LLMs and evaluating the performance of LLMs on the fault localization task.
After that, Widyasari et al. proposed FuseFL~\cite{DBLP:conf/wcre/WidyasariANS024}, which utilizes LLM to enhance the spectrum-based fault localization results.
Their research denotes that the LLM-powered explanation achieves high informativeness and clarity scores, statistically not significantly different from human explanations.
This work takes both the complete faulty code and the test case as input, however, in our scenario, the code scale is too large and the GUI test case is unavailable. 
Besides, Qin et al.~\cite{DBLP:journals/corr/abs-2403-16362} proposed AGENTFL for the project-level fault localization, which models the fault localization task as a three-step process, involving comprehension, navigation, and confirmation.
This tool outputs both the buggy location and the corresponding explanations, in which the explanations provide additional information to help developers make better fault localization.
Also, Kang et al.~\cite{kang2024quantitative} proposed AutoFL, which generates an explanation of the bug along with a suggested fault location using LLMs. 
Instead of finding bugs in particular code snippets, AutoFL focuses on performing fault localization in the entire software.
This is similar to our work, while we mainly concern the framework-based application crashing faults.
Besides, works~\cite{DBLP:journals/corr/abs-2403-16362, kang2024quantitative} mainly use LLM to finish the entire work.
Compared to them, which uses LLM to address the whole problem, our work uses LLM in a more lightweight mode.
To enhance the reliability of the explanation, instead of completely relying on the LLM, our approach tries to extract more summary knowledge with static analysis and use LLM to assist the final explanation generation.
}

\section{Conclusion}
\journal{ 
Post-release crash analysis is an inevitable challenge for application developers, demanding extensive efforts in debugging and fixing. 
This paper introduces a novel approach that combines static analysis and LLM to address the Android framework-specific fault localization and explanation problem.
It designed the ETS specification to encapsulate framework exception information and the CIS specification to represent the details of a buggy candidate. 
Leveraging static analysis, it tracks the buggy candidate within the application code and employs LLM to explain why the candidate may be erroneous.
In the aspect of fault localization, \ourtool{} outperforms the state-of-the-art tool with higher precision and fewer candidates. 
Regarding fault explanation, \ourtool{} achieves better results when compared to plain LLM and the naive results generated by pure static analysis.
}





\bibliographystyle{ieeetr}
\bibliography{bib/ref}







\end{document}